\documentclass{aa}
\usepackage{graphicx}
\usepackage{txfonts}

\begin{document}
\title{Detailed abundance analysis of globular clusters in the Local Group.}
\subtitle{NGC~147, NGC~6822, and Messier 33}

\author{
   S. S. Larsen
   \inst{1}
   \and
    J. Brodie
   \inst{2}
   \and
   A. Wasserman
   \inst{2}
   \and
    J. Strader
   \inst{3}
}

\institute{
  Department of Astrophysics/IMAPP,
              Radboud University, PO Box 9010, 6500 GL Nijmegen, The Netherlands\\
              \email{s.larsen@astro.ru.nl}
  \and
  UCO / Lick Observatory, 1156 High Street,
  University of California, Santa Cruz, CA 95064, USA
  \and
  Department of Physics and Astronomy, Michigan State University, East Lansing, Michigan 48824, USA
}

\date{Received September 7, 2017; accepted January 9, 2018}

\abstract
{Globular clusters (GCs) are emerging as powerful tracers of the chemical composition of extragalactic stellar populations.
}
{We present new abundance measurements for eleven GCs in the Local Group galaxies NGC~147, NGC~6822, and  Messier 33. These are combined with previously published observations of four GCs in the Fornax and WLM galaxies.
}
{The abundances were determined from analysis of integrated-light spectra, obtained with the HIRES spectrograph on the Keck~I telescope and with UVES on the VLT. We used our analysis technique that has been developed for this purpose and tested on Milky Way GCs.
}
{
We find that the clusters with $\mathrm{[Fe/H]}<-1.5$ are all $\alpha$-enhanced at about the same level as Milky Way GCs. Their Na abundances are also generally enhanced relative to Milky Way halo stars, suggesting that these extragalactic GCs resemble their Milky Way counterparts in containing significant fractions of Na-rich stars.
For $\mathrm{[Fe/H]}>-1.5$, the GCs in M33 are also $\alpha$-enhanced, while the GCs that belong to dwarfs (NGC~6822 SC7 and Fornax~4) have closer to Solar-scaled $\alpha$-element abundances.
The abundance patterns in SC7 are remarkably similar to those in the Galactic GC Ruprecht~106, including significantly sub-solar $\mathrm{[Na/Fe]}$ and $\mathrm{[Ni/Fe]}$ ratios.
In NGC~147, the GCs with $\mathrm{[Fe/H]}<-2.0$ account for about 6\% of the total luminosity of stars in the same metallicity range, a lower fraction than those previously found in the Fornax and WLM galaxies, but substantially higher than in the Milky Way halo.
}
{
At low metallicities, the abundance patterns suggest that GCs in the Milky Way, dwarf galaxies, and M33 experienced similar enrichment histories and/or processes. 
At higher metallicities, the lower levels of $\alpha$-enhancement in the GCs found in dwarf galaxies resemble the abundance patterns observed in field stars in nearby dwarfs. 
Constraining the presence of multiple populations in the GCs is complicated by the lack of information about detailed abundances in field stars of the corresponding metallicities. We suggest that correlations such as $\mathrm{[Na/Fe]}$ vs. $\mathrm{[Ni/Fe]}$ may prove useful for this purpose if an accuracy of $\sim0.1$ dex or better can be reached for the integrated-light measurements.
}


\maketitle

\section{Introduction}

Characterising and understanding the chemical composition of stellar populations is a fundamentally important problem in astrophysics. While elements heavier than hydrogen and helium only account for a couple of percent of the mass of stars like the Sun (and in some stars much less), these elements are of crucial interest. They were synthesised in stellar interiors, returned to the interstellar medium, and then incorporated in subsequent generations of stars, and therefore hold important clues to the evolutionary histories of galaxies. Of course, understanding the origin of the elements is also of interest beyond a purely astrophysical context. The striking similarity between the composition of stellar photospheres and the Earth's crust was noted long ago \citep{Payne1925}, although a detailed understanding of the synthesis of the elements in stellar interiors only started to emerge much later \citep[e.g.][]{Hoyle1954}.

\citet{Tinsley1979} discussed how abundance ratios can be affected by the different time scales on which various families of elements are produced (light, $\alpha$-, iron-peak, and neutron-capture elements). For example, elements that are produced predominantly in short-lived massive stars, such as oxygen, will tend to be overabundant compared to the iron-peak elements (produced in type Ia supernovae with longer-lived progenitors) in stars that formed early in the history of the Galaxy. Around the same time as Tinsley's pioneering theoretical work, observational evidence was accumulating that oxygen and calcium are indeed enhanced relative to iron in globular cluster giants \citep{Cohen1978,Pilachowski1980} and metal-poor halo stars \citep{Sneden1979,Luck1981}. 

It is impossible to review the extensive literature on more recent Galactic and extragalactic abundance work here. In the last two decades, a large number of studies have painted an increasingly detailed picture of the chemical composition of various Galactic components. Roughly speaking, the halo and bulge consist mainly of old, $\alpha$-enhanced populations, whereas young stellar populations in the disc tend to have abundance patterns similar to those seen in the Sun.  Fe-peak elements (such as Cr, Sc, Ni) mostly trace the Fe abundance, while the heavy (neutron-capture) elements (e.g. Ba, Y, La, Eu) show a large scatter and an increasing contribution from $r$-process dominated nucleosynthesis at low metallicities \citep{Edvardsson1993,Fulbright2000,Bensby2003,Reddy2003,Ishigaki2013,Bensby2014}. 
One interesting feature that has emerged from recent high-precision measurements is a bimodal distribution of $[\alpha/\mathrm{Fe}]$ ratios at metallicities $\mathrm{[Fe/H]}\lesssim-1$, with the less $\alpha$-enhanced stars having more extreme halo kinematics \citep{Nissen2010,Schuster2012,Bensby2014}. 
Many of the overall trends can be described fairly well by relatively simple models \citep{Matteucci1986,Matteucci1990,Pagel1995,Chiappini2001}, although chemo-dynamical models that capture more details are now also appearing \citep{Kobayashi2006,Kobayashi2011,Tissera2012}. 
Further details can be found in several excellent reviews \citep{McWilliam1998,Matteucci2001,Freeman2002,Helmi2008}.

A natural next step is to obtain abundance measurements for external galaxies and compare their chemical composition with our own Milky Way. This may help shed light on chemical enrichment in different environments and its implications for chemical evolution in the context of hierarchical galaxy assembly. For example, could the $\alpha$-rich and $\alpha$-poor halo populations described above be understood as stars formed ``in situ'' vs.\ accreted? In nearby dwarf galaxies, a considerable amount of information about detailed abundances is now available from observations of individual stars. Some differences with respect to the Galactic halo are evident: for example, the trend of $\alpha$-enhancement as a function of overall metallicity appears to be different in dwarf spheroidals than in the Milky Way, with the transition from significant $\alpha$-enhancement to more solar-like abundance ratios occurring at a lower metallicities in dwarf galaxies \citep{Tolstoy2009}. Other differences include increased $\mathrm{[Ba/Fe]}$ ratios and lower $\mathrm{[Na/Fe]}$ and $\mathrm{[Ni/Fe]}$ ratios in the dwarfs \citep{Cohen2009,Cohen2010,Letarte2010,Lemasle2014,Hendricks2016}. The reasons for these differences may include differences in the time scales for chemical enrichment, galactic winds, and variations in the stellar initial mass function \citep{Lanfranchi2006,Romano2013,McWilliam2013,Homma2015,Vincenzo2015}. 

In all but the closest of our neighbouring galaxies, individual stars are generally too faint for detailed abundance studies, especially for old stellar populations. 
Observations of early-type galaxies in integrated light have provided interesting constraints on $\alpha$-enrichment, which tends to increase with velocity dispersion
\citep{Worthey1992,Kuntschner2000,Trager2000,Thomas2005} and even on individual abundance ratios
\citep{Conroy2014}. 
However, a limitation of this type of work is that it only provides luminosity-weighted mean abundances, which are usually dominated by the more metal-rich populations. In order to access populations that only account for a minor fraction of the light, other methods are required. Here, \emph{globular clusters} (GCs) offer an attractive tracer of sub-populations within external galaxies as they are far brighter than individual red giants. In particular, the metal-poor components in galaxies usually account for only a small fraction of the stars, but often have large numbers of GCs associated with them.

A number of recent studies have used integrated-light spectroscopy at high resolution to measure chemical abundances of extragalactic GCs. In M31, as in the Milky Way, such measurements show that GCs generally have enhanced $\alpha$-element abundances, although some clusters show abundance patterns that more closely resemble those seen in dwarf galaxies \citep{Colucci2009,Colucci2014,Sakari2015}. A similar case in the Milky Way is Ruprecht~106, for which an accretion origin has been suggested \citep{Villanova2013}. In a pilot study we  used integrated-light spectroscopy to determine chemical abundances for three globular clusters (Fornax~3, Fornax~4, and Fornax~5) in the Fornax dwarf spheroidal galaxy. We confirmed that Fornax~3 and Fornax~5 are relatively metal-poor compared to typical Milky Way and M31 halo GCs (both having $\mathrm{[Fe/H]}<-2$), while Fornax~4 has $\mathrm{[Fe/H]}=-1.4$. The $[\alpha/\mathrm{Fe}]$ ratios ($\mathrm{[Ca/Fe]}$, $\mathrm{[Ti/Fe]}$) of Fornax~3 and Fornax~5 were found to be enhanced, although possibly slightly less so than in Milky Way GCs, while the $[\alpha/\mathrm{Fe}]$ ratio in Fornax~4 was only slightly elevated compared to Solar-scaled abundance patterns \citep[hereafter L12]{Larsen2012a}. This is in agreement with the trends observed in field stars in dwarf galaxies. Subsequently we applied our technique to the lone GC in the Wolf-Lundmark-Melotte (WLM) galaxy, which was also found to be metal-poor ($\mathrm{[Fe/H]}\approx-2$) and with chemical abundance patterns similar to those in Milky Way GCs \citep[L14]{Larsen2014}. We note that similar techniques have also been applied to \emph{young} star clusters in external star-forming galaxies \citep{Larsen2006b,Larsen2008a,Gazak2014,Lardo2015b,Cabrera-Ziri2016,Hernandez2017,Hernandez2017a} where they provide a welcome alternative to more traditional tracers of the chemical composition of young stellar populations (e.g., \ion{H}{ii} regions).

A unique aspect of GCs is the abundance spreads that affect the light elements within the clusters \citep{Gratton2012}. In nearly all Galactic GCs studied to date, about half of the stars have elevated abundances of certain elements (He, N, Na, and sometimes Al) while others are depleted (C, O, and sometimes Mg). Evidently, this needs to be kept in mind when using GCs as probes of stellar populations. On the other hand, integrated-light observations have the potential to constrain the occurrence of multiple populations in extragalactic environments. Indeed, the integrated-light abundances of elements like Na and Mg suggest that multiple populations are common also in extragalactic GCs \citep{Larsen2014,Colucci2014,Sakari2015}. 

Here we present new abundance measurements for GCs in three additional Local Group galaxies: NGC~147, NGC~6822, and M33. We use the same measurement technique that we have previously used for GCs in Fornax and WLM. Additional tests were described in \citet[][hereafter L17]{Larsen2017} where we compared integrated-light observations of seven Galactic GCs, spanning a broad range of metallicities, with literature data for individual stars in the clusters. In the following we briefly summarise previous work on chemical abundances in the three galaxies, concentrating on old stellar populations.

It is well-established that M33 has a metal-poor halo which is traced by red giant branch (RGB) stars \citep{Brooks2004} and RR Lyrae variables \citep{Sarajedini2006}, as well as GCs \citep{Christian1982,Sarajedini1998,Sarajedini2000}. These tracers all indicate an average metallicity of around $\mathrm{[Fe/H]}=-1.3$, but there is currently little information available about the detailed chemical composition of the M33 halo. From medium-dispersion spectroscopy of 15 GCs, \citet{Sharina2010} found a range of $[\alpha/\mathrm{Fe}]$ values between 0 and $+0.5$. A significant age spread among the M33 halo GCs has been suggested based on their horizontal branch (HB) morphologies \citep{Sarajedini1998}, and it would be interesting to investigate whether there is corresponding evidence of an extended formation history in the GC abundances. For example, one might expect the [$\alpha$/Fe] ratios to be closer to the Solar ratio if the clusters formed out of material that had time to become significantly enriched by type Ia SN ejecta. 

NGC~147 is a dE companion of M31 which is now known to host 10 old GCs, of which six have been discovered relatively recently \citep{Sharina2009,Veljanoski2013}. The metallicity distribution of the field stars in NGC~147 is quite broad with a peak at $\mathrm{[Fe/H]}=-0.5$ \citep{Ho2014}. In contrast, at least 7 out of the 10 GCs appear to have metallicities below $\mathrm{[Fe/H]}=-1.5$ \citep{Veljanoski2013}. NGC~147 thus exhibits the same offset between the GC and field star metallicity distributions that is observed in many other galaxies, including dwarf galaxies such as Fornax and WLM \citep{Larsen2012,Larsen2014,Lamers2017}. \citet{Vargas2014} estimated average $[\alpha/\mathrm{Fe}]$ ratios for RGB stars in NGC~147 and found them to be generally super-solar, although only one star had a metallicity below $\mathrm{[Fe/H]}=-1.5$. 

NGC~6822 (``Barnard's galaxy'') is a relatively isolated (within the Local Group) dwarf irregular galaxy, located at a distance of about 479 kpc \citep{Feast2012}. Its H{\sc ii} regions have long been known to be more metal-poor than those in the Milky Way \citep{Peimbert1970}. Modern studies based on various methods all agree on a metallicity close to $\mathrm{[Fe/H]}=-0.5$ for the young stellar population in NGC~6822: for H{\sc ii} regions, \citet{Lee2006a} found $\mathrm{[O/H]}=-0.55\pm0.10$, while \citet{Venn2001} found $\mathrm{[Fe/H]}=-0.49\pm0.22$ (blue supergiants) and \citet{Patrick2015} measured $\mathrm{[Fe/H]} = -0.52\pm0.21$ (red supergiants).
Information about older populations is more scarce, but \citet{Swan2016} found an average metallicity of $\langle\mathrm{[Fe/H]}\rangle = -0.84\pm0.04$ with a dispersion of 0.31 dex from \ion{Ca}{ii} IR triplet measurements of RGB stars. These authors also give a listing of other metallicity estimates. Until just a few years ago, NGC~6822 was thought to host only a single old GC, Hubble VII \citep{Hubble1925}. From integrated-light spectroscopy of Hubble VII, \citet{Colucci2011} found $\mathrm{[Fe/H]}=-1.61\pm0.02$, $\mathrm{[Ca/Fe]}=+0.01\pm0.07$ and $\mathrm{[Ba/Fe]}=+0.22\pm0.13$. Apart from this measurement, little is known about the detailed chemical composition of old stellar populations in NGC~6822.
In recent years, wide-field imaging has uncovered several previously uncatalogued GCs around NGC~6822 and the total now stands at 8 \citep{Hwang2011,Huxor2013}, several of which are compact and bright enough that good integrated-light spectra can be obtained in a modest amount of observing time with a 10 m class telescope.

\section{Observations}

\begin{table*}
\caption{Log of observations.}
\label{tab:obs}
\centering
\begin{tabular}{l c c c c c}
\hline\hline
Cluster & $V$ & Date & T(exp) & $v_\mathrm{hel}^a$ & $\sigma_\mathrm{1D}^b$ \\
            & mag &          & s            & km s$^{-1}$ & km s$^{-1}$ \\ \hline
M33 R12 & 16.5$^{1}$ & 24 Oct 1998 & $7\times1800$ s & $-218$ & 6.5 \\
M33 U49 & 16.0$^{1}$  & 24 Oct 1998 & $8\times1800$ s & $-150$ & 7.6 \\
M33 H38 & 17.4$^{1}$ &  25 Oct 1998 & $7\times1800$ s & $-241$ & 5.5 \\
M33 M9 & 17.1$^{1}$ & 25 Oct 1998 & $9\times1800$ s & $-249$ & 5.9 \\
NGC~147 HIII & 16.6$^2$ & 5 Oct 2015 & $4\times1800$ s  & $-197$ & 6.6 \\
NGC~147 HII & 18.1$^2$ & 25 Sep 2016 & $4\times1800$ s  & $-207$ & 2.5 \\
NGC~147 SD7 & 17.0$^2$ & 5 Oct 2015 & $4\times1800$ s & $-197$ & 5.3 \\
NGC~147 PA-1 & 17.0$^2$ & 5 Oct 2015 & 1800 s $+$ 865 s & $-221$ & 6.1 \\
& & 25 Sep 2016 & $2\times1800$ s \\
NGC~147 PA-2 & 17.4$^2$ & 25 Sep 2016 & $5\times1800$ s & $-221$ & 6.1 \\
NGC~6822 SC6 & 16.0$^3$ & 25 Sep 2016 & $2\times1800$ s & $-5$ & 8.7 \\
NGC~6822 SC7 & 15.4$^3$ & 5 Oct 2015 & 1800 s $+$ 1139 s & $-39$ & 9.2 \\
\hline
\end{tabular}
\tablefoot{
(a) Heliocentric radial velocity.
(b) One-dimensional velocity dispersion, corrected for instrumental resolution.
}
\tablebib{Integrated visual magnitudes:
(1)~\citet{Larsen2002b};
(2)~\citet{Veljanoski2013};
(3)~\citet{Veljanoski2015}.
}
\end{table*}

\begin{figure}
\centering
\includegraphics[width=\columnwidth]{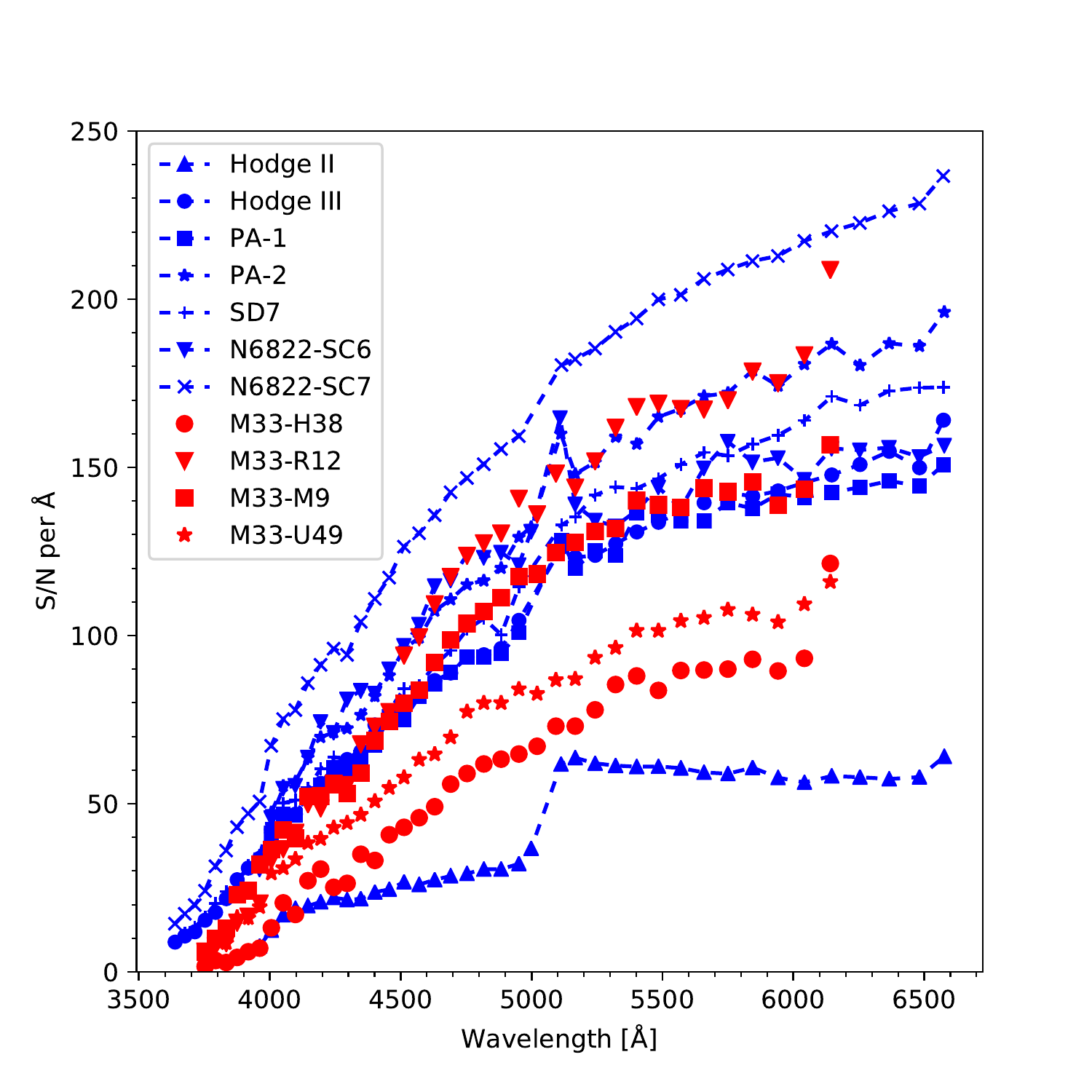}
\caption{\label{fig:s2n}Median signal-to-noise ratio per spectral order for the GCs in NGC~147, NGC~6822, and M33.}
\end{figure}

New observations of globular clusters in NGC~147 and NGC~6822 were obtained with the HIRES spectrograph \citep{Vogt1994} on the Keck~I telescope on 5 Oct 2015 and 25 Sep 2016. 
The observations on the night of 5 Oct 2015 were affected by clouds and high humidity (two nights had been scheduled for this run, of which the first was completely lost), whereas conditions during the Sep 2016 run were excellent. 
The clusters were selected from \citet{Veljanoski2013} and \citet{Veljanoski2015}, concentrating primarily on the brighter, more compact objects. A log of the observations is given in Table~\ref{tab:obs}. We used the C5 decker in HIRES, which provides a slit length of 7\arcsec\ and a slit width of $1\farcs148$, corresponding to a spectral resolving power of $R\approx37\, 000$. The detectors were binned by 2 pixels in the spatial direction, with a resulting spatial scale of $0\farcs24$ per (binned) pixel.

The clusters NGC~147-Hodge~II and NGC~6822-SC6 each has a relatively bright star at 8\arcsec--10\arcsec\ from the cluster centre. Although this is, in both cases, well beyond the half-length of the HIRES slit, we chose the slit position angle so as to avoid any potential light from these stars. For all other HIRES observations used in this paper, the slit was aligned vertically. This implies that the slit position angle on the sky changed during the exposures, typically by 5$^\circ$-20$^\circ$, depending on the zenith angle. The background level measured at the ends of the slit thus represents an average over many position angles and is naturally smoothed over any small-scale fluctuations that might be caused, for example, by fainter stars in the vicinity of the clusters.

The HIRES echellogram is recorded onto three separate detectors, which introduces small gaps in the wavelength coverage. The wavelength windows affected by these gaps can be chosen by adjusting the tilt of the echelle- and cross disperser gratings. We set the grating angles so as to have the gaps occur at 4990~\AA --5090~\AA\ and at 6600~\AA --6690~\AA, which minimised the loss of important spectral diagnostics. Of the features measured in L17, we only miss the \ion{Sc}{ii} line at 5031~\AA\ and several Ti lines near 5000~\AA . Apart from the gaps, the wavelength coverage is continuous from 3900~\AA\ to about 6300~\AA\, beyond which the ends of the echelle orders fall off the edges of the detectors.  The spectra extend to about 8100~\AA\ but in our analysis we concentrate on the spectral regions that we have used in our previous studies, i.e., 4200~\AA --6200~\AA .

We also include older HIRES observations of four GCs in M33. The spectra are the same as those used by \citet{Larsen2002b} to measure the internal velocity dispersions of the clusters and derive dynamical masses, and we refer to that paper for details about the observations and data reduction. These spectra were obtained on 24-25 October 1998. The detector then in use at HIRES covered the wavelength range 3730~\AA\--6170~\AA, distributed over 38 echelle orders with some gaps in the wavelength coverage between the orders at wavelengths longer than $\sim5000$~\AA. Since the original purpose of these observations was to measure velocity dispersions, a narrower slit ($0\farcs725$) was used, with a correspondingly higher resolving power of $R\approx54\,000$. The slit used for these observations had a length of $5\arcsec$. 
 The total integration times were 3.5--4.5 hrs, divided into 7--9 sub-exposures per cluster.

All HIRES spectra were reduced with the MAKEE (MAuna Kea Echelle Extraction) package\footnote{Available at http://www.astro.caltech.edu/\~{}tb/makee/} written by T.\ Barlow. 
MAKEE is designed to carry out all reduction steps with minimal user interaction, from bias subtraction and flat-fielding to tracing of the spectral orders, wavelength calibration, sky subtraction, and extraction of the final one-dimensional spectra using optimal weighting. Owing to the relatively short slit, definition of the sky regions involves a compromise between maximising the number of pixels used for the sky level determination and those used for extraction of the object spectrum. Also, especially for observations at high airmass, differential atmospheric refraction causes the clusters to appear off-center along the slit at bluer wavelengths. Hence, MAKEE determines the background regions individually for each echelle order based on the location of the centroid of the trace along the slit and the width of the object profile. The exact criteria controlling the sky subtraction and extraction are defined in a configuration file, where in most cases we used the default configuration file for HIRES provided with MAKEE. Here, the spacing between the object and sky regions is set to one pixel and the \emph{minimum} amount of sky on either side of the object is also one pixel. In practice, 2--5 binned pixels were typically used for sky determination on either side of the trace. In the absence of strong sky lines, MAKEE further averages the sky background over five adjacent pixels along the dispersion direction. The background subtracted from the object spectrum is then determined from a linear fit to the sky regions on either side of the trace (if the ratio of the sky areas is less than 2:1) or from a simple average of the sky pixels otherwise.
For NGC~147-Hodge~II, which is the faintest of our targets, we customised the configuration file to make MAKEE use at least five pixels on either side of the trace of the cluster spectra to determine the sky background level.
As a check we also reduced some of the spectra with the IDL HIRES Redux package written by J.\ Prochaska. The results were found to be very similar to those obtained with MAKEE.
 
The one-dimensional spectra extracted from each individual exposure were combined using a sigma-clipping algorithm to reject outlying pixels (e.g., due to cosmic ray hits), and the signal-to-noise (S/N) ratio of the combined spectra was estimated from the variance at each pixel.
In Fig.~\ref{fig:s2n} we show the S/N curves for our HIRES observations. Note that the S/N is given per \AA; the S/N per pixel is thus lower by a factor $\sqrt{\Delta \lambda/\AA}$ where $\Delta \lambda$ is the wavelength step per pixel, ranging from $\Delta \lambda = 0.018$~\AA\ at 4200~\AA\ to $\Delta\lambda=0.027$~\AA\ at 6200~\AA. For most clusters we reach a S/N$>$100 per \AA\ at 5000~\AA .

In addition to the HIRES observations, we include our previously published VLT/UVES spectra of GCs in the Fornax and WLM galaxies \citep{Larsen2012a,Larsen2014} and we refer to our previous papers for details on the observational strategy and data reduction. The Fornax GCs have half-light radii of several arcsec and for these clusters the UVES slit was scanned across the clusters in the north-south and east-west directions to sample the integrated light well. The science scans were bracketed by sky scans to facilitate sky subtraction. Basic reduction of the UVES data was carried out using the standard ESO pipeline (see L12 and L14 for details).
For homogeneity, we have reanalysed these data using the exact same procedures as the HIRES data, but the differences with respect to our previously published results are generally very minor.

\section{Analysis}

\begin{table}
\caption{Isochrone parameters and horizontal branches used for modelling of integrated-light spectra.}
\label{tab:iso}
\centering
\begin{tabular}{l c c c c}
\hline\hline
Object & Age & [Fe/H] & $[\alpha/\mathrm{Fe}]$ & HB from \\ \hline
NGC~147 HII & 13 & $-1.5$ & $+0.4$ & NGC~6254  \\
NGC~147 HIII &  13 & $-2.5$ & $+0.4$ & NGC~7078 \\
NGC~147 SD7 &  13 & $-2.0$ & $+0.4$ & NGC~6779 \\
NGC~147 PA-1 &  13 & $-2.3$ & $+0.4$ & NGC~7078 \\
NGC~147 PA-2 &  13 & $-2.1$ & $+0.4$ & NGC~6779 \\
NGC~6822 SC6 & 13 & $-1.8$ & $+0.4$ & NGC~6093 \\
NGC~6822 SC7 & 13 & $-1.3$ & 0.0 & NGC~362 \\
M33 H38 & 10 & $-1.2$ & $+0.4$ & NGC~362 \\
M33 M9 & 13 & $-1.8$ & $+0.4$ & NGC~6093 \\
M33 R12 & 10 & $-0.8$ & $+0.4$ & NGC~362 \\
M33 U49 & 10 & $-1.4$ & $+0.4$ & NGC~362 \\
WLM GC & 13 & $-2.0$ & $+0.2$ & NGC~6779 \\
Fornax 3 & \multicolumn{4}{c}{Empirical CMD} \\
Fornax 4 & \multicolumn{4}{c}{Empirical CMD} \\
Fornax 5 & \multicolumn{4}{c}{Empirical CMD} \\
\hline
\end{tabular}
\tablefoot{Isochrone ages (2nd column) are in Gyr. The last column indicates the names of the globular clusters from which horizontal branches were adopted.
}
\end{table}

Abundances were measured by means of our full spectral fitting technique, which has been described in detail and tested in our previous papers (L12, L14, L17). We continued to use the line list of \citet{Castelli2004} with the modifications discussed in L17.  We have updated the Mg oscillator strengths using data from the 18 Feb 2016 version of the Kurucz line list\footnote{http://kurucz.harvard.edu/linelists.html}, which are identical to those in the NIST database \citep{NIST}. As discussed in L17, these oscillator strengths agree better with those used by other recent studies. We have revised our procedure for selecting regions used to match the scaling of observed and model spectra. As noted in L14, using all available pixels when determining this scaling may lead to biased results, because model spectra are never a perfect match to real data. We have previously used the spectrum of Arcturus to identify ``continuum'' regions, free of strong lines, to be used for the scaling. Here we made use of the spectrum of 47~Tuc (NGC~104) analysed in L17, and selected regions where the NGC~104 spectrum and our best-fitting model spectrum agreed to within better than 2\%. With this modification, we used about 66\% of the pixels for overall scaling,  compared with about 32\% (in the range $4200 < \lambda/\AA < 6200$) with the selection based on the Arcturus spectrum.

To build our integrated-light model spectra, we used input Hertzsprung-Russell diagrams (HRDs) based on theoretical isochrones from the Dartmouth group for stellar evolutionary phases up to the tip of the RGB \citep{Dotter2007}. We generally assumed ages of 13 Gyr, while the metallicities and $\alpha$-enrichment of the isochrones were chosen to match the values derived from fits. This procedure typically converged after 1--2 iterations.  As in our previous work, the isochrones were populated according to a stellar mass function of the \citet{Salpeter1955} form, $\mathrm{d}N/\mathrm{d}M \propto M^{-2.35}$, down to a lower absolute magnitude limit of $M_V=+9$.
The isochrones were combined with empirical data for the horizontal branches (HBs) of Galactic GCs \citep{Sarajedini2007} with metallicities roughly matching those of the isochrones. As noted by \citet{Sarajedini1998}, some of the GCs in M33 have unusually red horizontal branches for their metallicities, which may be an indication that they are somewhat younger than their Galactic counterparts, although the mass-to-light ratios are similar to those of Galactic GCs \citep{Larsen2002b}. For these clusters we assumed ages of 10 Gyr and we used the (red) HB of the globular cluster NGC~362. 
For the globular clusters in the Fornax dSph we used the same input HRDs, based on HST colour-magnitude diagrams, as in L12.
 Table~\ref{tab:iso} summarises the isochrone parameters and HB data used for each cluster. 
 
\begin{figure}
   \centering
   \includegraphics[width=\columnwidth]{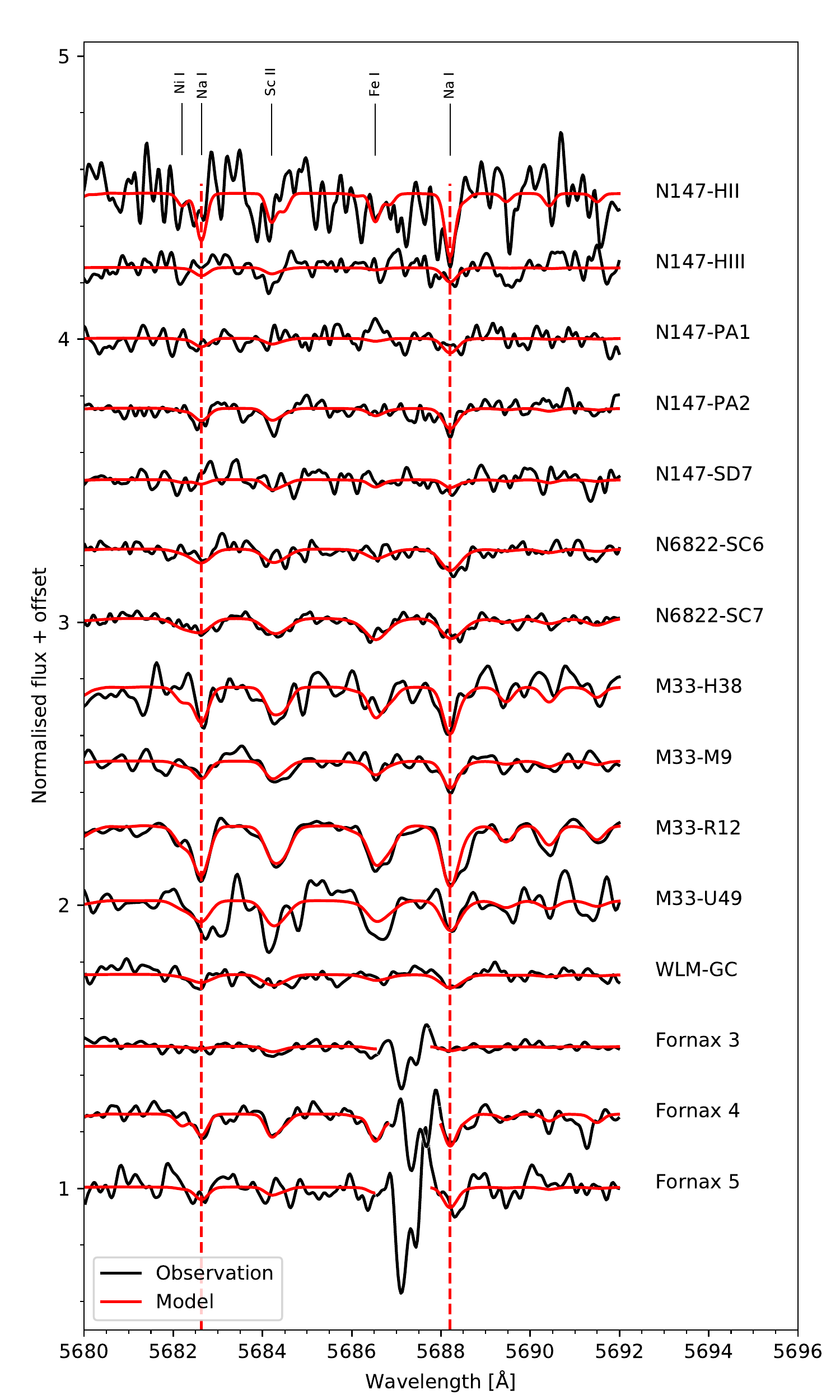}
      \caption{\label{fig:fitna57}Fits to the \ion{Na}{i} doublet at 5683/5688~\AA\ (red dashed lines). Note that the bad pixels near 5687~\AA\ in the Fornax spectra have been excluded from the fits.
       Both models and data have been smoothed using a Gaussian kernel with $\sigma=2$ pixels.}
   \end{figure}

\begin{figure}
   \centering
   \includegraphics[width=\columnwidth]{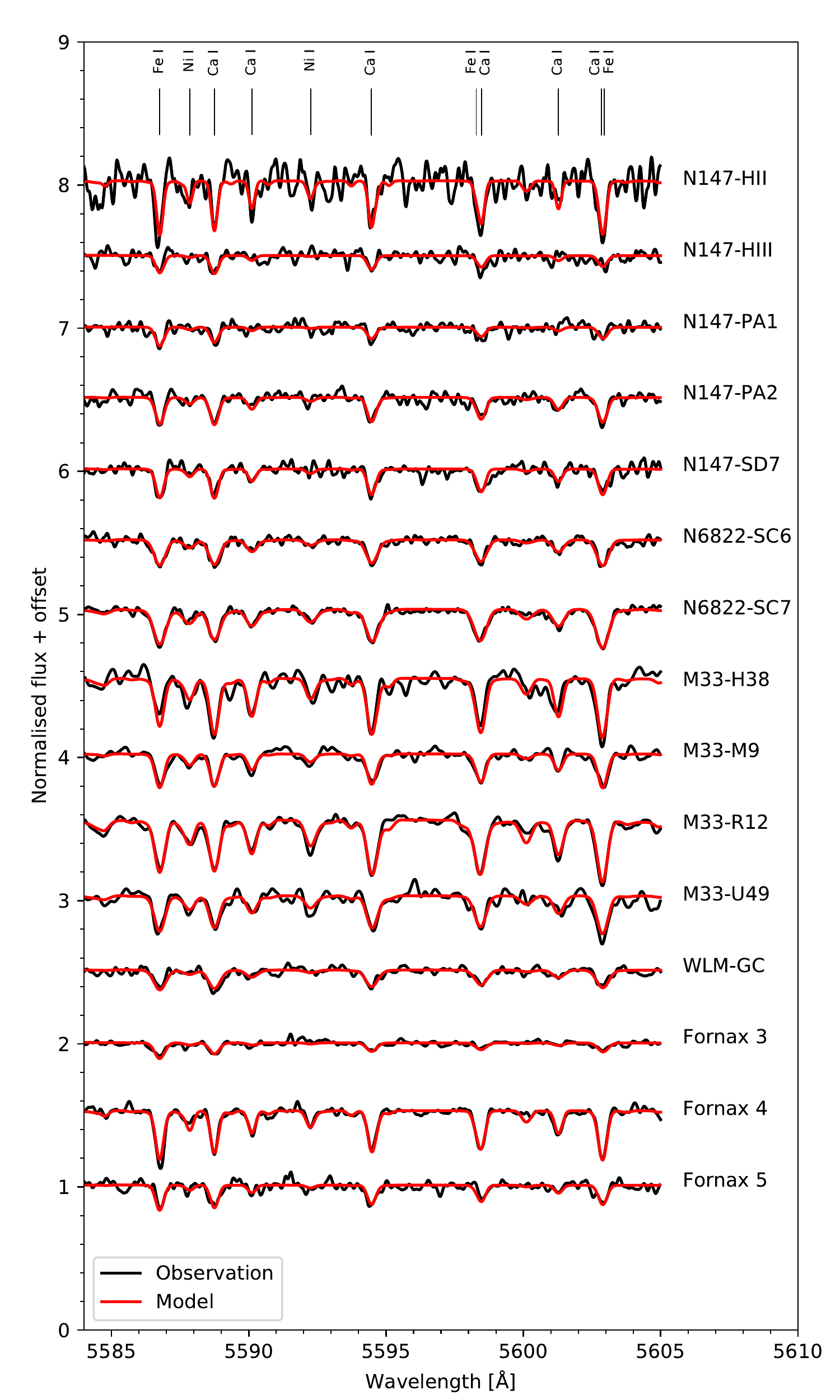}
      \caption{\label{fig:fitca56}Fits to the spectral region near 5600 \AA . Both models and data have been smoothed using a Gaussian kernel with $\sigma=2$ pixels.}
   \end{figure}

In general, abundances of Na, Mg, Ca, Sc, Ti, Cr, Mn, and Ba were determined using the same spectral bins as in our previous work. However, since the HIRES reduction pipelines do not merge the echelle orders to a single spectrum, we determined the Fe abundances by fitting each echelle order separately, rather than in 200 \AA\ bins as was done for our UVES observations. A new addition is that we also included 14 bins containing Ni lines. In a few cases, the edges of the spectral bins had to be adjusted in order to stay within one echelle order. For the M33 GC spectra, a few features fell in the gaps between echelle orders, allowing us to measure only three Mg lines instead of the five we measured for the other clusters. For the Fornax spectra we now include measurements of Na, which was omitted in L12 because of a block of bad pixels near the 5683/5688 \AA\ doublet. However, the 6154/6161~\AA\ lines are strong enough to measure in the Fornax~4 spectrum, and for this spectrum we found the two \ion{Na}{i} doublets to give $\mathrm{[Na/Fe]} = -0.177\pm0.045$ (5683/5688~\AA) and $\mathrm{[Na/Fe]} = -0.258\pm0.120$ (6154/6161~\AA), i.e., consistent within the measurement errors. This is also in agreement with our analysis in L17, which showed that the spectral synthesis technique performs well even for the weaker and more blended \ion{Na}{i} lines at 6154/6161 \AA . 
Hence, we now feel sufficiently confident about these measurements that we choose to quote Na abundances also for the Fornax GCs. Fig.~\ref{fig:fitna57} shows the fits to the \ion{Na}{i} 5683/5688~\AA\ doublet for all clusters studied here and it can be seen that the bad pixels in the Fornax spectra (which are masked out in the fits) are located just bluewards of the 5688~\AA\ line.  Fig.~\ref{fig:fitca56} shows additional example fits to the spectral region near 5600~\AA, which contains a number of \ion{Ca}{i} lines. 

We include in Table~\ref{tab:obs} the heliocentric radial velocities ($v_\mathrm{hel}$) and line-of-sight velocity dispersions ($\sigma_\mathrm{1D}$) obtained as part of our analysis. For the velocity dispersions, we have subtracted the instrumental resolution in quadrature 
($\sigma_\mathrm{inst} = 2.4$ km s$^{-1}$ for the 1998 observations,  $\sigma_\mathrm{inst} = 3.4$ km~s$^{-1}$ otherwise). The errors on the radial velocities are dominated by systematics, but are probably accurate to within $\sim1$ km~s$^{-1}$ (L17). For Hodge~III, \citet{Sharina2009} quoted a velocity of $-118\pm30$ km~s$^{-1}$, which differs substantially from the systemic velocity of NGC~147 \citep[$-193.1\pm0.8$ km s$^{-1}$;][]{Geha2010}. Our measurement ($-197$ km s$^{-1}$) is much closer to the velocity of NGC~147. For the other clusters in NGC~147 and NGC~6822, our new measurements agree well with previous literature data \citep{Sharina2009,Veljanoski2013,Veljanoski2015}.
For U49 and R12 in M33, the radial velocities derived from our analysis agree well (to within about 1 km s$^{-1}$) with those found by \citet{Larsen2002b} from cross-correlation with the star HD~1918. However, for H38 and M9 the velocities found here are 47 km s$^{-1}$ lower (more negative) than those found by \citet{Larsen2002b}. This is probably due to a misidentification of the template star: U49 and R12 were observed on the first night of the 1998 observing run, apparently together with the correct star. The two other clusters were observed the following night, and inspection of the original FITS headers for the template star observations reveals that the telescope was pointed at another star about 1$\arcmin$ from the position of HD~1918. We have verified by cross-correlating the two template spectra that their radial velocities do indeed differ by 47~km~s$^{-1}$. Apparently the star observed on the second night was not HD 1918 but another star. Our new radial velocities for R12, U49, and H38 agree with those measured by \citet{Chandar2002} within the errors quoted in that study (10--20 km s$^{-1}$), while our velocity for M9 is about 36~km~s$^{-1}$ more negative.
The velocity dispersions for the M33 GCs in Table~\ref{tab:obs} agree with the ``direct fit'' values derived by \citet{Larsen2002b} within about 0.5 km s$^{-1}$, with no significant differences between clusters observed on the first and second night. This suggests that the template star misidentification does not significantly influence the results in that paper (apart from the radial velocities).

\subsection{Age and horizontal branch morphology}

\begin{table}
\caption{Sensitivity to SSP modelling assumptions for M33-U49.}
\label{tab:sens}
\centering
\begin{tabular}{l c c}
\hline\hline
   & $t=13$ Gyr & NGC 5272 HB \\ \hline
$\Delta$[Fe/H]  & $-0.063$ & $+0.058$ \\
$\Delta$[Na/Fe] & $+0.034$ & $-0.036$ \\
$\Delta$[Mg/Fe] & $-0.046$ & $+0.125$ \\
$\Delta$[Ca/Fe] & $+0.011$ & $-0.070$ \\
$\Delta$[Sc/Fe] & $+0.034$ & $-0.025$ \\
$\Delta$[Ti/Fe] & $+0.008$ & $+0.052$ \\
$\Delta$[Cr/Fe] & $+0.019$ & $-0.026$ \\
$\Delta$[Mn/Fe] & $+0.013$ & $-0.001$ \\
$\Delta$[Ni/Fe] & $+0.004$ & $+0.008$ \\
$\Delta$[Ba/Fe] & $-0.008$ & $+0.061$ \\
\hline
\end{tabular}
\tablefoot{Differences are listed with respect to the ``standard'' abundances derived for $t=10$ Gyr and the NGC~362 horizontal branch.}
\end{table}

In our previous papers (L12, L14, L17) we have assessed many of the uncertainties and systematics related to particular choices made in the modelling. Given that the horizontal branch morphology is uncertain for many of the clusters studied here, we carried out an additional set of fits for one of the clusters in M33 (U49), using the blue horizontal branch of the Galactic GC NGC~5272 instead of the red horizontal branch of NGC~362. The resulting changes in the abundance ratios are listed in Table~\ref{tab:sens}. To assess the effect of age uncertainties, we also list abundance changes for an assumed age of 13 Gyr instead of 10 Gyr. The differences are generally small (less than 0.1 dex). Changing the HB morphology has the largest effect on the $\mathrm{[Mg/Fe]}$ and $\mathrm{[Ca/Fe]}$ ratios (which increase by 0.13 dex and decrease by 0.07 dex, respectively) and also increases the iron abundance by 0.06 dex. Changing the age decreases the iron abundance by 0.06 dex, but hardly affects most abundance ratios. 

\section{Results}
\label{sec:results}

\begin{table*}
\small
\caption{Abundance measurements.}
\label{tab:abun}
\centering
\begin{tabular}{l r r r r r r r r r r r}
\hline\hline
 & $\mathrm{[Fe/H]}$ & $\mathrm{[Na/Fe]}$ & $\mathrm{[Mg/Fe]}$ & $\mathrm{[Ca/Fe]}$ & $\mathrm{[Sc/Fe]}$ & 
 $\mathrm{[Ti/Fe]}$    &     $\mathrm{[Cr/Fe]}$    &     $\mathrm{[Mn/Fe]}$    &     $\mathrm{[Ni/Fe]}$ & $\mathrm{[Ba/Fe]}$ \\
 & rms$_w$ (N) & rms$_w$ (N) & rms $_w$(N) & rms$_w$ (N) & rms$_w$ (N) & rms$_w$ (N) & rms$_w$ (N) & rms$_w$ (N) & rms$_w$ (N) & rms$_w$ (N) \\
\hline
NGC 147 HII  & $-1.541$  & $+0.398$  & $+0.254$  & $+0.260$  & $+0.147$  & $+0.316$  & $-0.111$  & $-0.164$  & $+0.248$  & $-0.203$  \\
 & 0.279 (24)  & 0.085$^a$ (1)  & 0.206 (4)  & 0.215 (6)  & 0.269 (5)  & 0.153 (6)  & 0.230 (8)  & 0.269 (2)  & 0.342 (11)  & 0.427 (3) \\
NGC 147 HIII  & $-2.480$  & $+0.648$  & $+0.190$  & $+0.447$  & $+0.243$  & $+0.509$  & $-0.198$  & $+0.200$  & $-0.027$  & $-0.255$  \\
 & 0.154 (25)  & 0.091$^a$ (1)  & 0.150 (4)  & 0.136 (6)  & 0.407 (5)  & 0.198 (8)  & 0.288 (7)  & 0.299 (2)  & 0.455 (5)  & 0.065 (4) \\
NGC 147 PA-1  & $-2.370$  & $+0.552$  & $+0.298$  & $+0.227$  & $+0.023$  & $+0.372$  & $+0.002$  & $+0.032$  & $-0.276$  & $+0.172$  \\
 & 0.133 (22)  & 0.146$^a$ (1)  & 0.177 (5)  & 0.222 (8)  & 0.389 (5)  & 0.159 (9)  & 0.353 (10)  & 0.080 (1)  & 0.300 (4)  & 0.137 (4) \\
NGC 147 PA-2  & $-2.011$  & $+0.343$  & $+0.395$  & $+0.390$  & $+0.136$  & $+0.384$  & $+0.017$  & $-0.244$  & $+0.075$  & $+0.408$  \\
 & 0.093 (27)  & 0.085$^a$ (1)  & 0.161 (5)  & 0.134 (8)  & 0.232 (6)  & 0.129 (9)  & 0.182 (10)  & 0.098 (2)  & 0.226 (11)  & 0.064 (4) \\
NGC 147 SD7  & $-2.057$  & $-0.192$  & $+0.532$  & $+0.364$  & $+0.002$  & $+0.432$  & $+0.048$  & $-0.409$  & $+0.195$  & $+0.432$  \\
 & 0.111 (26)  & 0.166$^a$ (1)  & 0.396 (5)  & 0.137 (8)  & 0.303 (6)  & 0.146 (10)  & 0.257 (11)  & 0.027 (2)  & 0.333 (12)  & 0.341 (4) \\
NGC 6822 SC6  & $-1.822$  & $+0.284$  & $+0.295$  & $+0.228$  & $+0.110$  & $+0.290$  & $-0.049$  & $-0.426$  & $-0.058$  & $+0.408$  \\
 & 0.078 (27)  & 0.090$^a$ (1)  & 0.240 (6)  & 0.170 (8)  & 0.158 (7)  & 0.091 (9)  & 0.123 (11)  & 0.068 (2)  & 0.203 (13)  & 0.107 (4) \\
NGC 6822 SC7  & $-1.271$  & $-0.366$  & $-0.180$  & $+0.042$  & $-0.318$  & $+0.013$  & $-0.112$  & $-0.388$  & $-0.137$  & $+0.283$  \\
 & 0.092 (26)  & 0.070$^a$ (1)  & 0.208 (5)  & 0.141 (8)  & 0.118 (7)  & 0.128 (10)  & 0.098 (11)  & 0.020 (2)  & 0.236 (14)  & 0.044 (4) \\
M33 H38  & $-1.176$  & $-0.020$  & $+0.016$  & $+0.293$  & $-0.024$  & $+0.367$  & $+0.016$  & $-0.571$  & $+0.189$  & $+0.657$  \\
 & 0.215 (23)  & 0.096$^a$ (1)  & 0.409 (2)  & 0.355 (8)  & 0.108 (3)  & 0.383 (12)  & 0.310 (8)  & 0.178 (2)  & 0.346 (12)  & 0.129 (4) \\
M33 M9  & $-1.739$  & $+0.207$  & $-0.030$  & $+0.263$  & $+0.105$  & $+0.351$  & $-0.037$  & $-0.378$  & $+0.069$  & $+0.588$  \\
 & 0.101 (26)  & 0.081$^a$ (1)  & 0.165 (3)  & 0.177 (9)  & 0.235 (6)  & 0.154 (12)  & 0.110 (9)  & 0.115 (2)  & 0.268 (12)  & 0.130 (3) \\
M33 R12  & $-0.942$  & $+0.098$  & $+0.206$  & $+0.285$  & $+0.087$  & $+0.261$  & $+0.026$  & $-0.215$  & $+0.068$  & $+0.395$  \\
 & 0.079 (26)  & 0.041$^a$ (1)  & 0.083 (3)  & 0.144 (9)  & 0.141 (6)  & 0.146 (12)  & 0.157 (11)  & 0.067 (2)  & 0.192 (13)  & 0.166 (4) \\
M33 U49  & $-1.446$  & $+0.057$  & $+0.266$  & $+0.219$  & $+0.257$  & $+0.380$  & $+0.109$  & $-0.023$  & $+0.145$  & $+0.544$  \\
 & 0.133 (26)  & 0.115$^a$ (1)  & 0.154 (3)  & 0.337 (8)  & 0.230 (5)  & 0.246 (11)  & 0.290 (10)  & 0.137 (2)  & 0.379 (10)  & 0.407 (3) \\
WLM GC  & $-1.947$  & $+0.217$  & $+0.057$  & $+0.236$  & $+0.213$  & $+0.177$  & $-0.113$  & $-0.474$  & $-0.059$  & $+0.099$  \\
 & 0.088 (10)  & 0.141$^a$ (1)  & 0.129 (5)  & 0.121 (6)  & 0.181 (6)  & 0.164 (9)  & 0.185 (6)  & 0.083 (2)  & 0.196 (12)  & 0.169 (4) \\
Fornax 3  & $-2.353$  & $+0.013$  & $-0.014$  & $+0.239$  & $+0.185$  & $+0.311$  & $-0.189$  & $-0.466$  & $-0.016$  & $+0.554$  \\
 & 0.066 (10)  & 0.171$^a$ (1)  & 0.163 (5)  & 0.157 (8)  & 0.183 (6)  & 0.101 (9)  & 0.095 (6)  & 0.133 (2)  & 0.259 (13)  & 0.089 (3) \\
Fornax 4  & $-1.374$  & $-0.177$  & $-0.014$  & $+0.073$  & $-0.077$  & $+0.124$  & $-0.089$  & $-0.365$  & $-0.179$  & $+0.272$  \\
 & 0.052 (10)  & 0.045$^a$ (1)  & 0.189 (5)  & 0.130 (8)  & 0.095 (6)  & 0.094 (9)  & 0.093 (6)  & 0.053 (2)  & 0.155 (15)  & 0.060 (4) \\
Fornax 5  & $-2.125$  & $+0.484$  & $+0.205$  & $+0.273$  & $+0.012$  & $+0.267$  & $+0.059$  & $-0.367$  & $+0.049$  & $-0.005$  \\
 & 0.076 (10)  & 0.096$^a$ (1)  & 0.154 (5)  & 0.174 (8)  & 0.263 (6)  & 0.088 (9)  & 0.181 (6)  & 0.066 (1)  & 0.223 (14)  & 0.127 (4) \\
\hline
\end{tabular}
\tablefoot{
For each cluster, the first line gives the average abundance ratio and the second line gives the weighted rms and number of individual measurements.
(a): For $\mathrm{[Na/Fe]}$ we list the errors on the individual measurements of the doublet at 5683/5688 \AA .
}
\end{table*}

The individual abundance measurements for each spectral bin are given in Tables~\ref{tab:abN147HII}-\ref{tab:abF5}. 
Bins for which the spectral fitting procedure failed to converge to a value within $-1 < \mathrm{[X/Fe]} < +1$ (typically because of low S/N) are marked as ``$\ldots$'' and in some cases only upper limits could be determined. The average abundance measurements are listed in Table~\ref{tab:abun}. For each cluster, we give the weighted mean of the measurements for each individual bin, as well as the weighted r.m.s. (rms$_w$, calculated as in L17) and number of measurements $N$. The weights were assigned as described in L17, i.e. a ``floor'' of 0.01 dex has been added in quadrature to the formal errors on the individual measurements. Bins for which the spectral fitting procedure failed to converge, or only upper limits were obtained, have been excluded. For Na, Table~\ref{tab:abun}  only uses the values based on the 5683/5688~\AA\ doublet, although Tables~\ref{tab:abN147HII}-\ref{tab:abF5} also list measurements of the weaker 6154/6161~\AA\ lines (that typically have large errors). As found in L17, the two \ion{Na}{i} doublets generally yield consistent abundances within the errors.

Our new measurements for the Fornax and WLM clusters generally differ little with respect to our previously published values. 
The largest differences are for Ba, with our new analysis yielding $\mathrm{[Ba/Fe]}$ values that are lower by 0.2--0.3 dex. As discussed in L17, this is due to a combination of updated $\log gf$ values and the fact that we now assume an $r$-process dominated isotopic mixture for Ba (effectively making hyperfine structure more important). The $\mathrm{[Mg/Fe]}$ ratios have increased systematically by on average 0.04 dex, which can again be attributed to updated $\log gf$ values (L17). This brings the Mg abundances into closer agreement with the other alpha-elements (Ca, Ti), although $\mathrm{[Mg/Fe]}$  is still significantly lower than the $\mathrm{[Ca/Fe]}$ and $\mathrm{[Ti/Fe]}$ ratios for a few clusters. We discuss these clusters in more detail below (Sect.~\ref{sec:alphafe}). For other abundances ratios, the systematic differences between our previous and current analyses are less than 0.02 dex.

\subsection{Metallicities: comparisons with previous work}

\begin{table}
\caption{Literature metallicity determinations.}
\label{tab:prevz}
\centering
\begin{tabular}{l c c c}
\hline\hline
Object & $\mathrm{[Fe/H]}$ & Method & Ref. \\ \hline
NGC~147 HII 
    & $-1.2\pm0.4$ & a & SD2009 \\
    & $-1.7\pm0.4$ & c & V2013  \\
NGC~147 HIII & $-2.1\pm0.3$ & c & V2013  \\
     & $-2.5\pm0.25$ & b & DM88 \\
NGC~147 SD7 
   & $-1.6\pm0.2$ & a & SD2009 \\
   & $-1.8\pm0.3$ & c & V2013  \\      
NGC~147 PA-1 & $-2.0\pm0.3$ & c & V2013  \\
NGC~147 PA-2 & $-1.9\pm0.3$ & c & V2013  \\
NGC~6822 SC6 & $-1.40^{+0.07}_{-0.04}$ & d & V2015 \\
 & $-1.7\pm0.3$ & c & V2015 \\
NGC~6822 SC7 & $-0.61^{+0.03}_{-0.04}$ & d & V2015 \\
 & $-1.0\pm0.4$ & c & V2015 \\
M33 H38 & $-1.10\pm0.10$ & e & S98  \\
M33 M9 &  $-1.64\pm0.28$ & e & S98 \\
M33 R12 &  $-1.19\pm0.24$ & e & S98 \\
 & $-0.73\pm0.01$ & a & S2010 \\
M33 U49 & $-1.64\pm0.20$ & e & S98 \\
 & $-1.70\pm0.53$ & b & BH91 \\
\hline
\end{tabular}
\tablefoot{
(a): Spectral fitting;
(b): Spectroscopic line indices; 
(c): Integrated broad-band colours \citep[via relation from][]{KisslerPatig2002}; 
(d): Integrated broad-band colours \citep[via comparison with SSP models based on][]{Bressan2012};
(e): Red giant branch slope
}
\tablebib{
DM88:~\citet{DaCosta1988};
BH91:~\citet{Brodie1991};
SD2009:~\citet{Sharina2009};
V2013:~\citet{Veljanoski2013};
V2015:~\citet{Veljanoski2015};
S98:~\citet{Sarajedini1998}; 
S2010:~\citet{Sharina2010}
}
\end{table}

In Table~\ref{tab:prevz} we list  previous estimates of the cluster metallicities based on various techniques. 
In most cases, these previous studies have not measured $\mathrm{[Fe/H]}$ directly, but have estimated it based on line index measurements, broad-band colours, or RGB slope, as indicated in the Table.
For Hodge~II and SD7 in NGC~147 we quote the $\mathrm{[Fe/H]}$ values from \citet{Sharina2009}, which were obtained using continuum shape fitting. 

In NGC~147, the overall ranking of the cluster metallicities according to the literature studies agrees well with our measurements: Hodge~II is consistently found to be the most metal-rich of the clusters, followed by PA-2 and SD7, while Hodge~III and  PA-1 are the most metal-poor. Indeed, our measurement of $\mathrm{[Fe/H]}=-2.5$ for Hodge~III places this cluster among the most metal-poor GCs known, with a metallicity similar to or slightly below those of the most metal-poor GCs in the Milky Way such as M15, M30, M92, and NGC~5053 \citep{Harris1996}, and comparable to Fornax~1, the most metal-poor GC in the Fornax dwarf galaxy \citep{Letarte2006,Larsen2014a}. For SC6 and SC7 in NGC~6822 there are no previous spectroscopic metallicity determinations, but \citet{Veljanoski2015} estimated metallicities from broad-band colours using both the empirical colour-metallicity relation of \citet{KisslerPatig2002} and comparison with simple stellar population (SSP) models \citep[based on PARSEC isochrones;][]{Bressan2012}. Our
spectroscopic metallicity determinations agree better with those based on the empirical colour-metallicity relation, whereas the metallicities based on SSP model comparisons appear to overestimated.

For the M33 clusters, our overlap with previous studies is limited. For U49, our metallicity determination agrees with the previous spectroscopic metallicity estimate by \citet{Brodie1991} within the (large) uncertainty on their measurement. For R12, our new metallicity is 0.22 dex lower than the spectroscopic estimate of $\mathrm{[Fe/H]}=-0.73\pm0.01$ of \citet{Sharina2010}, although the 0.01 dex uncertainty on that measurement appears somewhat optimistic (the same paper quotes an overall metallicity of $\mathrm{[Z/H]} = -0.6\pm0.2$). Our measurements generally agree with those based on the RGB slope from \citet{Sarajedini1998}, within the 0.1--0.3 dex uncertainties on those estimates. 
From full spectral fitting, \citet{Beasley2015} found metallicities of $\mathrm{[Z/H]} = -1.15\pm0.25$ (M33-H38), $-1.74\pm0.16$ (M33-M9), $-0.86\pm0.17$ (M33-R12), and $-1.33\pm0.16$ (M33-U49). Converting these metallicities to iron abundances using the relation $\mathrm{[Fe/H]} = \mathrm{[Z/H]} - 0.75 \times \mathrm{[Mg/Fe]}$ \citep{Vazdekis2015} and our measured $\mathrm{[Mg/Fe]}$ values, this yields
$\mathrm{[Fe/H]} = -1.16\pm0.40$, $-1.72\pm0.18$, $-1.02\pm0.18$, and $-1.53\pm0.18$ for the four clusters. While it should be borne in mind that the conversion from $\mathrm{[Z/H]}$ to $\mathrm{[Fe/H]}$ does not account for the detailed abundance patterns, these iron abundances are in excellent agreement with those in Table~\ref{tab:abun}.

\subsection{Metallicity distributions of globular clusters and field stars  in NGC 147}

\begin{figure}
\centering
\includegraphics[width=\columnwidth]{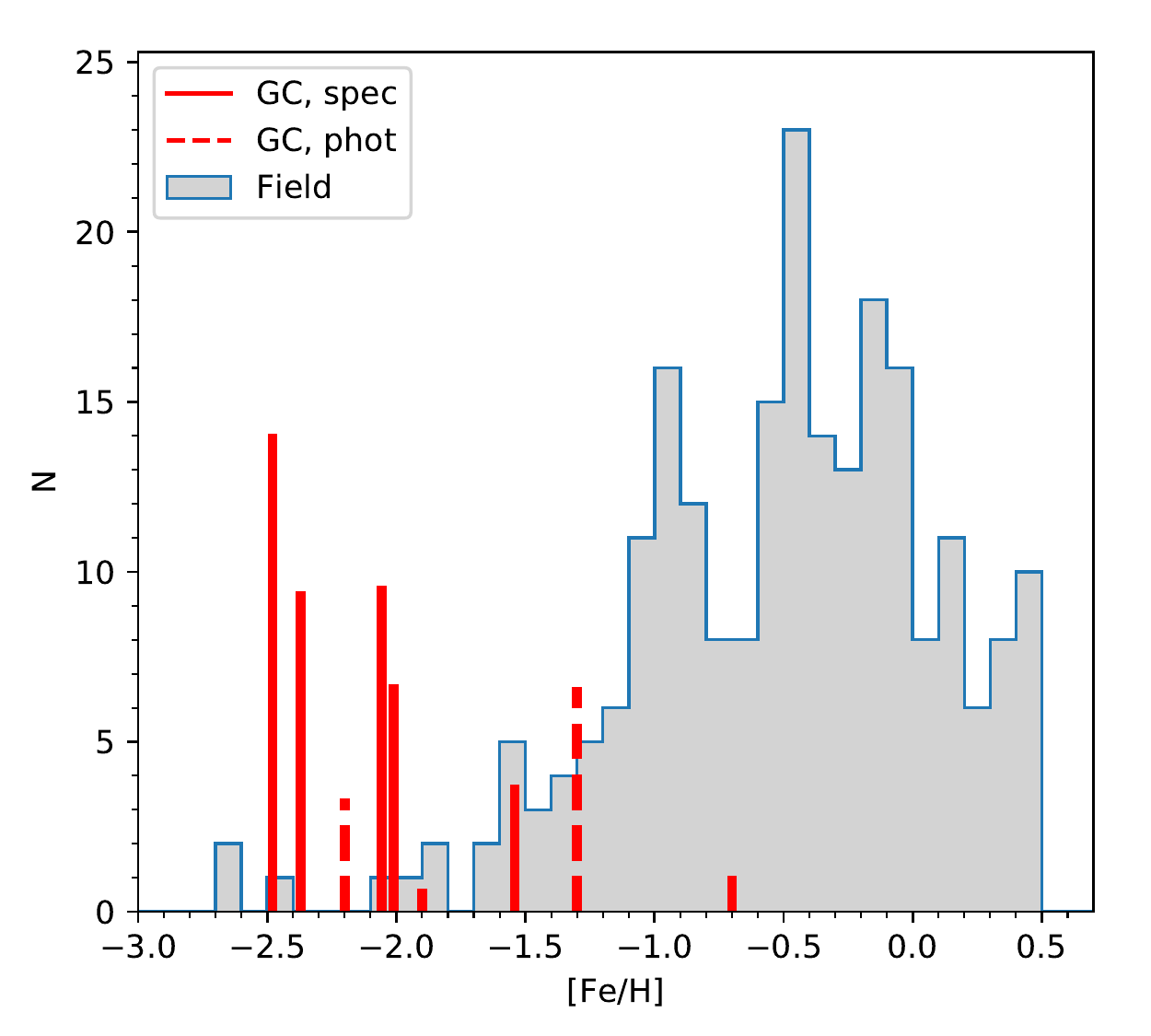}
\caption{\label{fig:fecmp_n147}Metallicity distributions of field stars \citep{Ho2014} and GCs in NGC~147. For the GCs, the bars have been scaled according to the $V$-band luminosity of each cluster. Solid bars: this work. Dashed bars: \citet{Veljanoski2013}.}
\end{figure}

It is well-known that the metallicity distributions of field stars and GCs in galaxies often differ substantially, with GCs being preferentially associated with the more metal-poor stellar populations \citep{Forte1981,Forbes2001,Harris2002,Harris2007}. This phenomenon is particularly extreme in some dwarf galaxies such as Fornax and WLM (L12, L14), where the GCs account for a significant fraction ($\sim1/5$--1/4) of the metal-poor stars. 

Our new metallicity determinations for GCs in NGC~147 allow us to carry out a similar comparison for this galaxy.
The metallicity distribution for field stars in NGC~147 has been determined by \citet{Ho2014} via measurements of the \ion{Ca}{ii} infrared triplet. In Fig.~\ref{fig:fecmp_n147} we plot a histogram of their metallicity measurements together with vertical bars indicating the GC metallicities. The solid bars indicate our spectroscopic measurements whereas the dashed bars are the estimates by \citet{Veljanoski2013}, which are based on broad-band colours. The lengths of the bars are scaled according to the luminosities of the GCs. Although we do not have spectroscopic measurements for all clusters, those for which we do have such measurements tend to be the brighter ones, which account for most of the mass and luminosity of the GC system.

The usual difference between GC and field star metallicity distributions is evident from Fig.~\ref{fig:fecmp_n147}. \citet{Ho2014} quote an average metallicity of $\langle\mathrm{[Fe/H]}\rangle = -0.51$ for the field stars, whereas most of the GCs have metallicities 1--2 dex below this value. In fact, there are so few metal-poor field stars that it is difficult to reliably estimate what fraction of the low-metallicity stars belong to GCs. Only five stars (out of a total of 230) have $\mathrm{[Fe/H]}<-2$, corresponding to about 2.2\% of the measured stars. 

The outer regions of NGC~147 are perturbed due to tidal interactions with M31, which makes it challenging to estimate the total luminosity.  \citet{McConnachie2012} lists an absolute magnitude of $M_V=-14.6$, whereas \citet{Harris2013} give a value of $M_V = -15.46\pm0.3$. From a S{\'e}rsic profile fit, \citet{Crnojevic2014} found $M_V=-16.5$.
If we simply scale the luminosity estimated by \citet{Crnojevic2014} by 2.2\%, we then find that the metal-poor stars have a total magnitude of $M_V= -12.34$. The total luminosity of the GCs with $\mathrm{[Fe/H]}<-2.0$ corresponds to $M_V=-9.42$, or about 6.4\% of the luminosity of the metal-poor stars. This is a lower fraction than those previously found for the Fornax dSph and WLM, but still substantially higher than for the Milky Way halo where the fraction is about 2\% (e.g. L14).
If we instead adopt the absolute magnitude from \citet{McConnachie2012} then the fraction increases to 25\%. 

A more detailed calculation should account for the fact that translating the number fraction of bright metal-poor RGB stars to a total mass- or luminosity fraction requires knowledge of the star formation history.  However, using isochrones from \citet{Dotter2007}, we find that for a magnitude cut of $M_I=-3$ (roughly corresponding to the magnitude limit in Ho et al.) the number of RGB stars per unit total luminosity is about independent of metallicity and age for ages $>$ 4 Gyr. From deep HST photometry, \citet{Geha2015} found that NGC~147 has an extended star formation history with a significant intermediate-age (5--7 Gyr) population. Hence, the number fraction of metal-poor stars may be used as a reasonable proxy for the contribution of metal-poor stars to the total luminosity of NGC~147. \citet{Ho2014} argue that biases in their spectroscopic metallicity distribution due to colour biases are likely small. Their spectra cover the full radial extent of NGC~147 and there are, in any case, no obvious gradients in the metallicity distribution \citep{Crnojevic2014}. Clearly,  a dominant uncertainty in the analysis carried out here is the small number of metal-poor stars.

\subsection{Individual abundances}
\label{sec:abun}

\subsubsection{Alpha-elements: Mg, Ca, Ti}
\label{sec:alphafe}

   \begin{figure}
   \centering
   \includegraphics[width=\columnwidth]{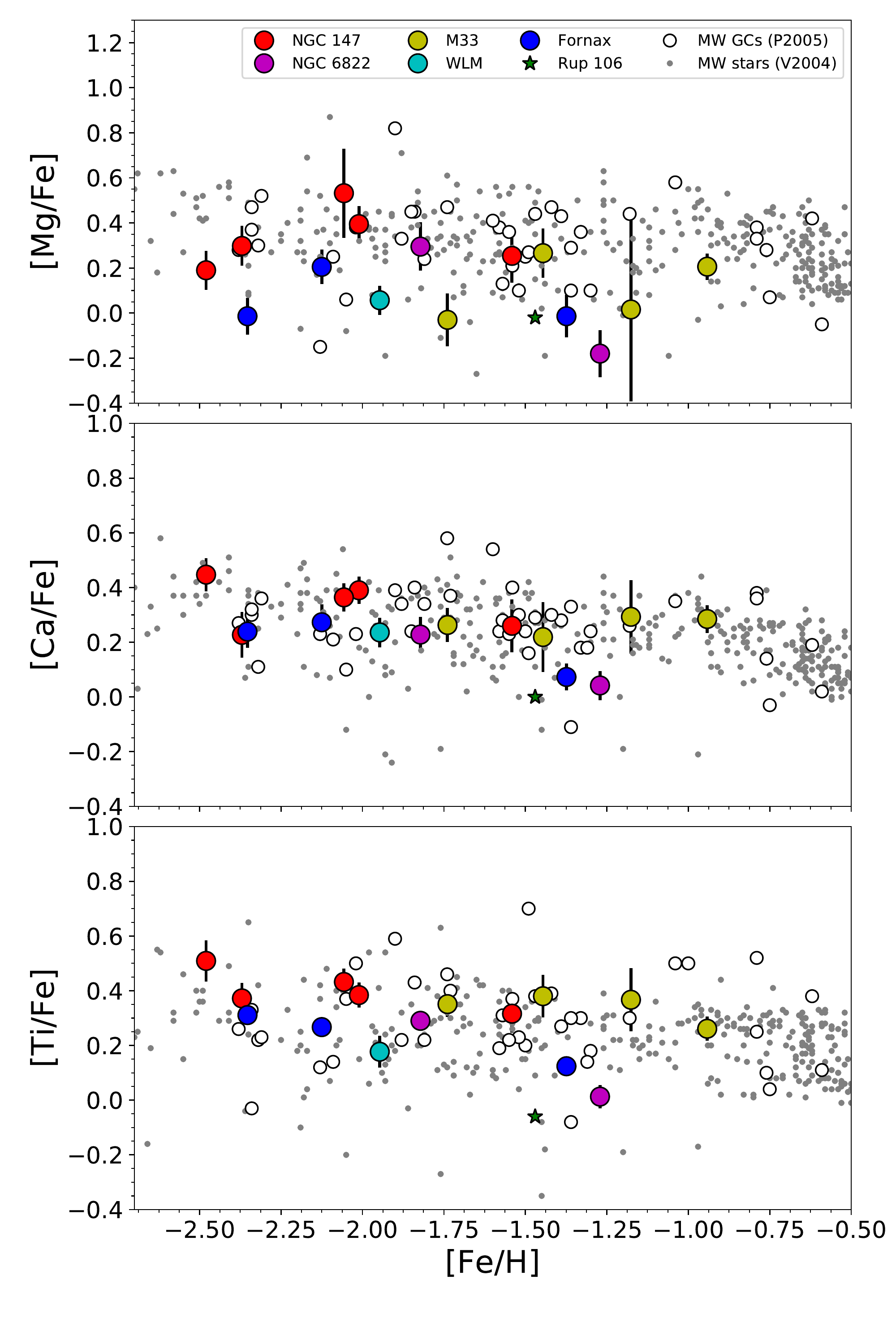}
      \caption{
         \label{fig:alphafe}Alpha-element abundances versus metallicity. Symbols are colour-coded according to host galaxy as indicated in the legend. Also included are data for Milky Way globular clusters \citep[open circles,][]{Pritzl2005} and field stars \citep[grey dots,][]{Venn2004}. Data for the Galactic GC Rup 106 are shown with a green star \citep{Villanova2013}.
         }
   \end{figure}

   \begin{figure}
   \centering
   \includegraphics[width=\columnwidth]{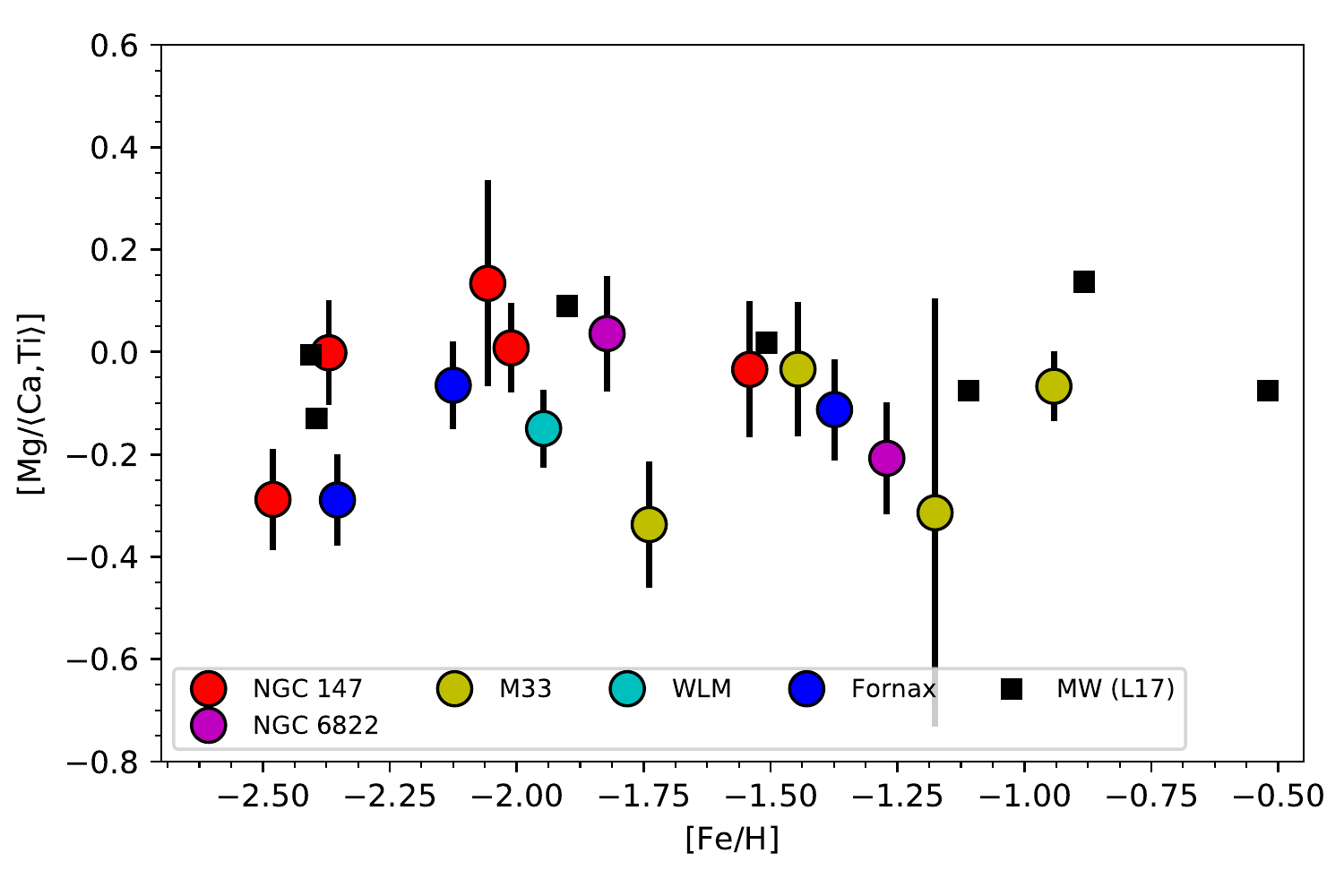}
      \caption{
         \label{fig:mgafe}$[\mathrm{Mg}/\langle\mathrm{Ca,Ti}\rangle]$ for our integrated-light measurements. Also included are our data for Milky Way GCs (L17).
         }
   \end{figure}

In Fig.~\ref{fig:alphafe} we plot our measurements of the $\alpha$-element abundances ($\mathrm{[Mg/Fe]}$, $\mathrm{[Ca/Fe]}$, and $\mathrm{[Ti/Fe]}$) as a function of $\mathrm{[Fe/H]}$. 
For comparison we also include compilations of data for Milky Way GCs \citep{Pritzl2005} and field stars \citep{Venn2004}. At low metallicities ($\mathrm{[Fe/H]}\la-1.5$), our integrated-light $\mathrm{[Ca/Fe]}$ and $\mathrm{[Ti/Fe]}$ ratios for extragalactic GCs are similar to those observed in Milky Way GCs and field stars, being enhanced relative to Solar-scaled abundance patterns by about 0.3 dex (i.e., about a factor of two). At higher metallicities we see a hint of a dichotomy between the abundance patterns in the M33 GCs and those in the dwarf galaxies (Fornax~4 and NGC~6822 SC7): the metal-rich M33 GCs remain $\alpha$-enhanced at about the same level as their more metal-poor counterparts, whereas the GCs in the dwarfs have lower $\mathrm{[Ca/Fe]}$ and $\mathrm{[Ti/Fe]}$ ratios. 
A lower degree of $\alpha$-enhancement in the more metal-rich GCs in the dwarfs is in accordance with the trend observed among field stars in nearby dwarf galaxies (Fornax, Sculptor, Sagittarius, Carina), in which $[\alpha/\mathrm{Fe}]$ decreases towards solar-scaled values between $\mathrm{[Fe/H]}=-2$ and $\mathrm{[Fe/H]}=-1$ \citep{Tolstoy2009,Letarte2010,Hendricks2016}. The GC Fornax~4 is thus consistent with the trends seen in the Fornax field stars, and our measurements for NGC~6822 SC7 provide the first evidence that a similar pattern exists for old stellar populations in NGC~6822.

In field stars, $\mathrm{[Mg/Fe]}$ tends to follow the same abundance patterns as other $\alpha$-elements, at least within the metallicity range probed here \citep{McWilliam1997,Tolstoy2009}. Departures from this behaviour have previously been noted for integrated-light analyses of extragalactic GCs in M31 \citep{Colucci2009,Colucci2014}, with $\mathrm{[Mg/Fe]}$ being lower than other alpha-element ratios. Similar offsets have been seen in our own work (L12, L14). Such offsets might potentially be caused by internal Mg abundance spreads in the clusters, although Mg spreads that would be large enough to cause the observed effects are rare in Galactic GCs \citep{Carretta2009}.
While our use of updated Mg oscillator strengths tends to reduce the difference, we see from Fig.~\ref{fig:alphafe} that there may still be a Mg deficiency for some clusters. To better assess the difference between Mg and the other alpha-elements, we plot the ratio $[\mathrm{Mg}/\langle\mathrm{Ca,Ti}\rangle]$ for our integrated-light measurements in Fig.~\ref{fig:mgafe}. For comparison we also include our integrated-light measurements for seven Galactic GCs (Table B.2 in L17). Of the 15 extragalactic GCs, three (Hodge~III, M33-M9, and Fornax~3) have relatively low ($<-0.2$) $[\mathrm{Mg}/\langle\mathrm{Ca,Ti}\rangle]$ ratios, two more may also be depleted, although the difference with respect to the Milky Way sample is less significant given the errors (NGC~6822-SC7 and M33-H38), whereas the remaining clusters do not differ significantly from the Milky Way clusters. 

It is worth asking how robust the apparently sub-solar $[\mathrm{Mg}/\langle\mathrm{Ca,Ti}\rangle]$ ratios are. 
We first note that all clusters in Fig.~\ref{fig:mgafe} have been analysed using the same methodology, so that systematic effects and biases should be as minimal as possible. A closer look at the individual $\mathrm{[Mg/Fe]}$ measurements for Hodge~III, M33-M9, and Fornax~3 (Tables~\ref{tab:abN147HIII}, \ref{tab:abM33M9}, and \ref{tab:abF3}) shows that the low average Mg abundances are not driven by a single outlying measurement. For Hodge~III, all four individual measurements give $\mathrm{[Mg/Fe]}$ ratios below the average $\mathrm{[Ca/Fe]}$ and $\mathrm{[Ti/Fe]}$ ratios, and the weighted rms (rms$_w=0.15$ dex) is comparable to the individual errors (0.10-0.21 dex), which suggests that most of the line-to-line scatter can be accounted for by the random measurement uncertainties. For M33-M9, rms$_w$ is again similar to the errors on the individual measurements, all of which indicate low $\mathrm{[Mg/Fe]}$ values, although in this case only three lines could be measured. For Fornax~3, the weighted rms is somewhat larger than the individual errors. The  \ion{Mg}{i} line at 5711~\AA\ gives $\mathrm{[Mg/Fe]}=+0.34\pm0.11$, which is consistent with the average $\mathrm{[Ca/Fe]}$ and $\mathrm{[Ti/Fe]}$ ratios, but the other four Mg lines all yield $\mathrm{[Mg/Fe]} < +0.1$. For M33-H38, the uncertainty on the $\mathrm{[Mg/Fe]}$ measurement (based on only two lines) is so large that no meaningful statement can be made, whereas the moderately low $[\mathrm{Mg}/\langle\mathrm{Ca,Ti}\rangle]$ ratio for NGC~6822-SC7 is indeed, to some extent, driven by one line (\ion{Mg}{i} at 4352~\AA).
In summary, we only consider the sub-solar $[\mathrm{Mg}/\langle\mathrm{Ca,Ti}\rangle]$ ratios in the three GCs Hodge~III, M33-M9, and Fornax~3 as relatively robust.

\subsubsection{Sodium}
\label{sec:nafe}

   \begin{figure}
   \centering
   \includegraphics[width=\columnwidth]{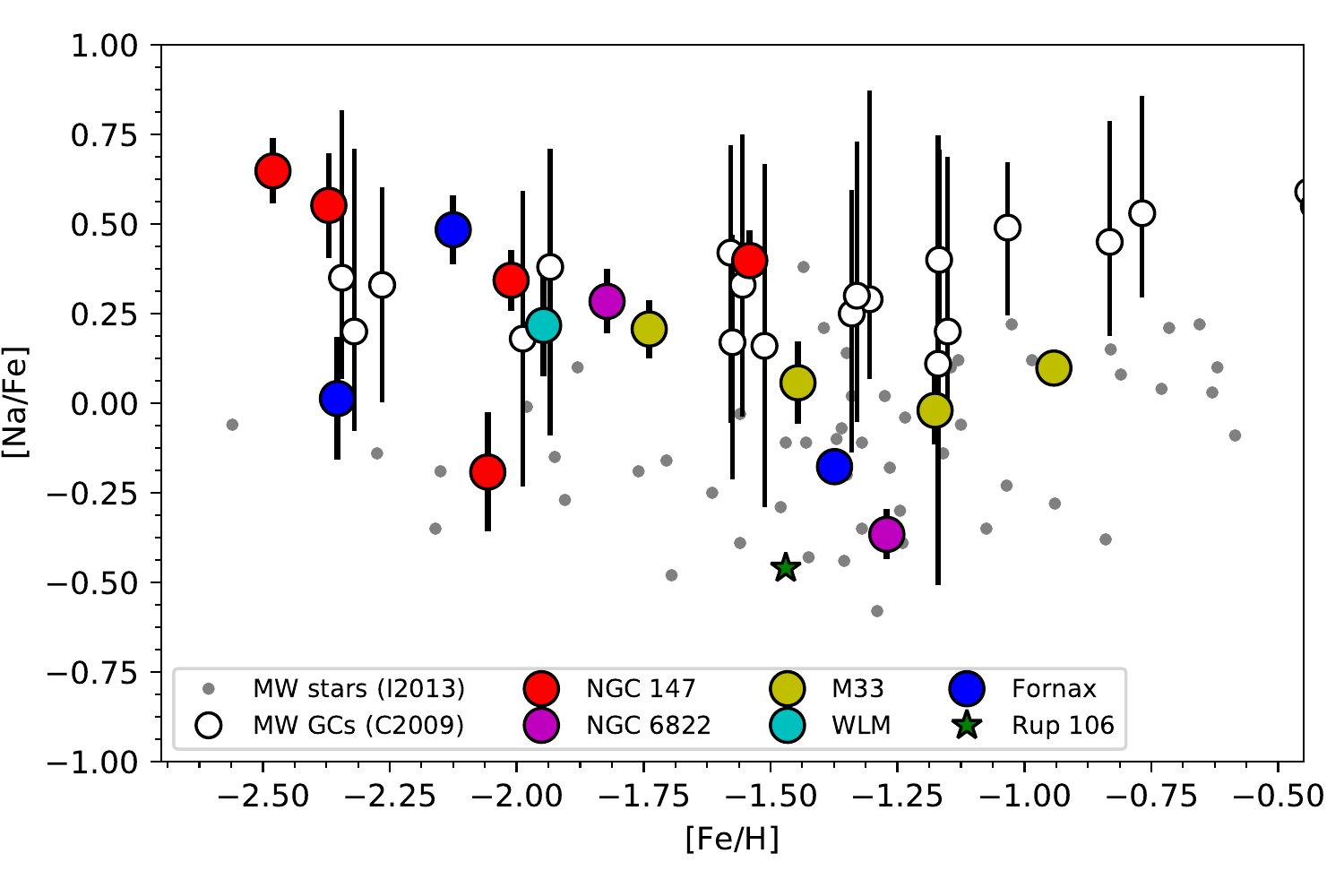}
      \caption{
         \label{fig:nafe}Sodium abundances versus metallicity. Also included are data for Milky Way globular clusters \citep{Carretta2009} and field stars \citep[grey dots,][]{Ishigaki2013}.
         }
   \end{figure}

Figure~\ref{fig:nafe} shows our integrated-light measurements of $\mathrm{[Na/Fe]}$, together with data for Milky Way field stars \citep{Ishigaki2013} and globular clusters \citep{Carretta2009}. 
For the Carretta et al.\ data, the vertical bars show the range of $\mathrm{[Na/Fe]}$ values observed within the clusters and the open circles indicate the average values. The lower end of the range of $\mathrm{[Na/Fe]}$ values typically coincides with the abundances observed in the field, while GCs contain stars in which the $\mathrm{[Na/Fe]}$ ratios are enhanced by up to about 0.5 dex. Consequently, the average $\mathrm{[Na/Fe]}$ ratios in GCs are higher compared to those seen in the field. 

For most of the extragalactic GCs,  the integrated-light $\mathrm{[Na/Fe]}$ ratios resemble the average values in Milky Way GCs in being elevated, but again there are a few exceptions. 
We find relatively low $\mathrm{[Na/Fe]}$ ratios for
NGC~147 SD7 and Fornax~3, although the Na lines are weak in these metal-poor clusters (cf.\ Fig.~\ref{fig:fitna57}) and the uncertainties on the measurements are relatively large. We note that the 6154/6161~\AA\ lines give a higher sodium abundance of $\mathrm{[Na/Fe]} = +0.43\pm0.19$ for NGC~147 SD7, while no abundance could be determined from these lines for Fornax~3. However, we remind the reader that the Na abundance for Fornax~3 might be affected by the bad pixels near the 5688~\AA\ line. Potentially more significant are the low $\mathrm{[Na/Fe]}$ ratios of the two more metal-rich GCs, NGC~6822 SC7 and, to a lesser extent, Fornax~4. In both cases, the 6154/6161~\AA\ lines confirm sub-solar $\mathrm{[Na/Fe]}$ ratios, albeit with a large uncertainty for NGC~6822 SC7.
This may simply reflect the general tendency for Na to be relatively under-abundant in dwarf galaxies, with typical abundance ratios of  $\mathrm{[Na/Fe]} \simeq -0.5$  at intermediate metallicities \citep{Shetrone2003,Tolstoy2009}. 
 No information on the detailed abundances of halo field stars is available for NGC~6822. However, we recall that NGC~6822 SC7 also has relatively low $\alpha$-element abundance ratios, comparable to those of Fornax~4.
 Hence, our measurements of $\mathrm{[Na/Fe]}$ may still leave room for an enhanced mean Na abundance relative to the (unknown) ``baseline'' level in the field stars. 
 
The M33 GCs may also have slightly lower $\mathrm{[Na/Fe]}$ ratios than Milky Way GCs at the same metallicity, but interpretation is again complicated by our lack of knowledge about the Na abundances in M33 halo field stars. Additionally, we note that the absolute scale of the Na abundances is somewhat uncertain, as our integrated-light measurements do not currently include corrections for non-LTE effects, which are often applied to measurements of individual stars. These corrections can reach 0.2-0.3 dex, but their magnitude (and even the sign) depends on the gravity, temperature, and composition of the star \citep{Gratton1999}. In L17 we found that our integrated-light $\mathrm{[Na/Fe]}$ ratios for Galactic GCs were systematically slightly lower (by 0.08--0.14 dex, depending on the comparison sample) than average values derived from analysis of individual stars, which suggests that our integrated-light Na abundances may be underestimated by approximately that amount. 

\subsubsection{Iron-peak elements: Sc, Cr, Mn, Ni}

   \begin{figure}
   \centering
   \includegraphics[width=\columnwidth]{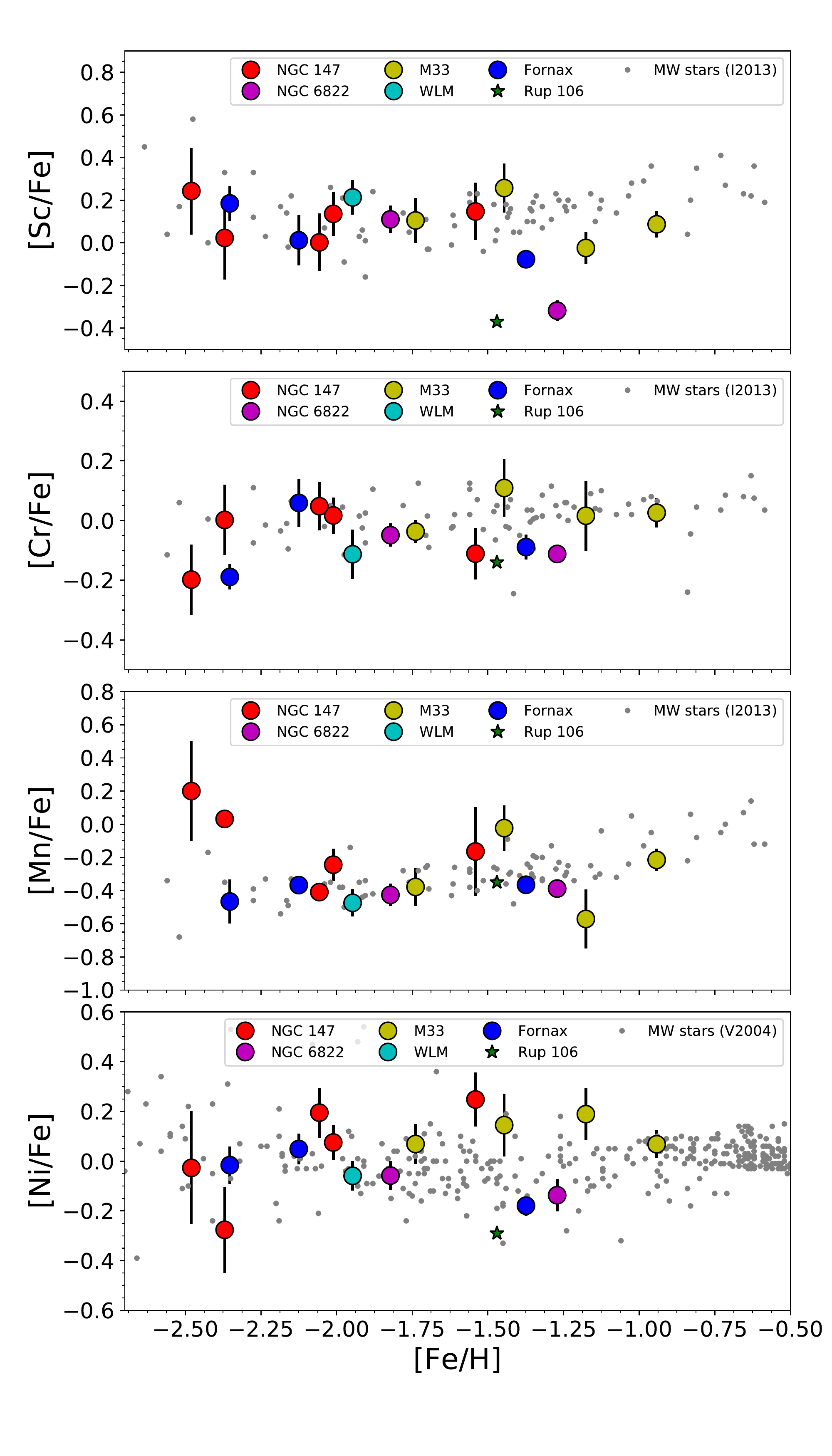}
      \caption{
         \label{fig:fepeak}Iron-peak element abundances versus metallicity. Also included are data for Milky Way field stars \citep[grey dots,][]{Ishigaki2013,Venn2004}.
         }
   \end{figure}

In Fig.~\ref{fig:fepeak} we show the iron-peak element abundances (Sc, Cr, Mn, Ni). These elements are usually unaffected by the presence of multiple populations in GCs and thus display similar behaviour in GCs and field stars.  Our integrated-light abundance ratios for these elements are fairly similar to those observed in the Milky Way. Manganese follows the well-known trend of increasing from $\mathrm{[Mn/Fe]}\simeq-0.5$ at low metallicities to $\mathrm{[Mn/Fe]}\simeq 0$ at higher metallicities \citep{Nissen2000}, although this trend may be mainly an artefact of non-LTE effects \citep{Bergemann2008,Battistini2015}. 
Perhaps significantly, the two most metal-poor clusters (both in NGC~147) have high $\mathrm{[Mn/Fe]}$ ratios.

NGC~6822~SC7 and Fornax~4 are, once again, outliers in some of these relations, with relatively low $\mathrm{[Sc/Fe]}$ and $\mathrm{[Ni/Fe]}$ ratios. In this sense, they resemble the Galactic globular cluster Ruprecht~106, which is marked with an asterisk \citep{Villanova2013}. The low Sc abundances in these clusters are consistent with the correlation between Sc and the $\alpha$-elements observed in Galactic field stars \citep{Nissen2000,Ishigaki2013}. 

\subsubsection{Barium}

   \begin{figure}
   \centering
   \includegraphics[width=\columnwidth]{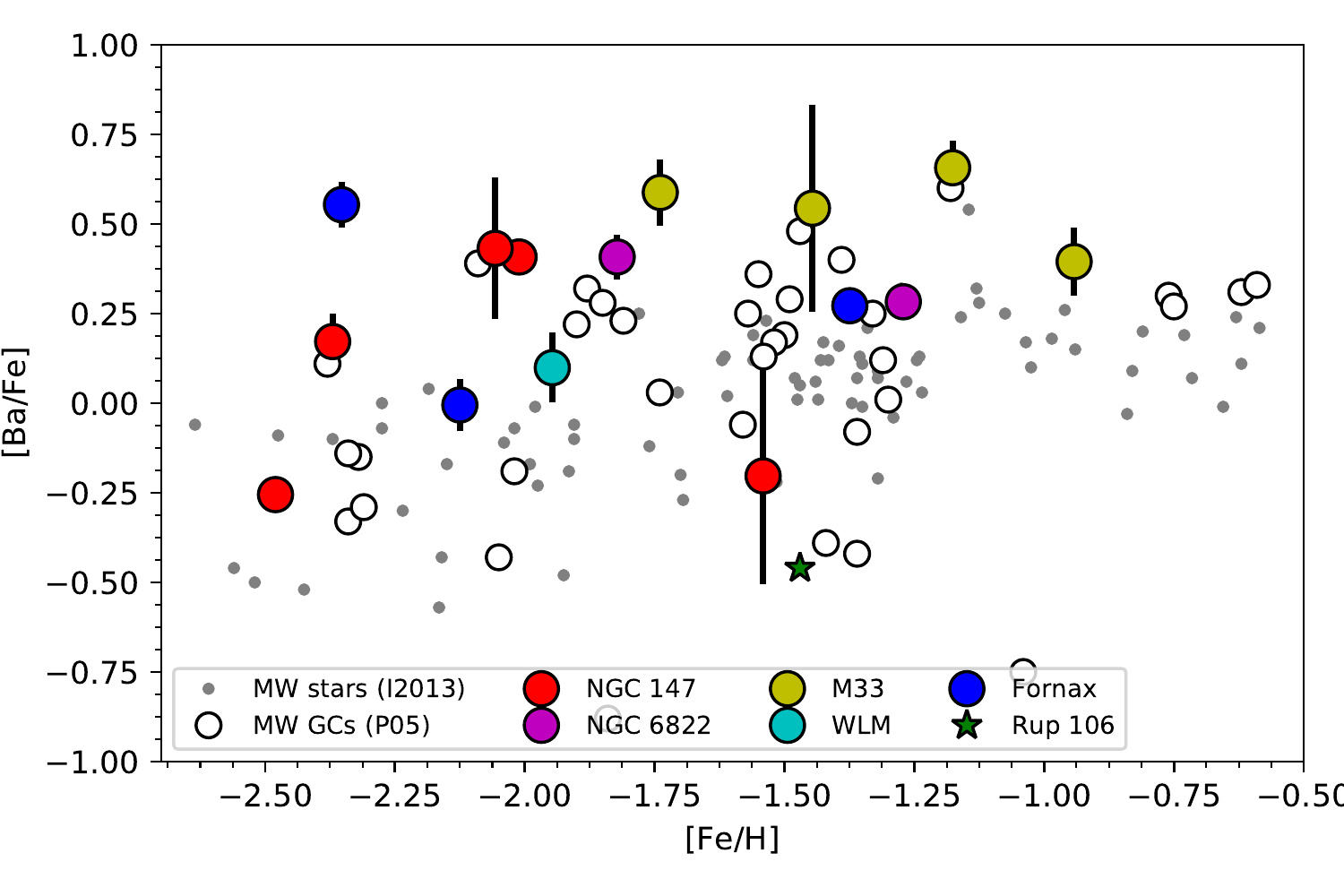}
      \caption{
         \label{fig:bafe}Barium abundances versus metallicity. Also included are data for Milky Way globular clusters \citep{Pritzl2005} and field stars \citep[grey dots,][]{Ishigaki2013}.
         }
   \end{figure}

The $\mathrm{[Ba/Fe]}$ ratios show substantial scatter (Fig.~\ref{fig:bafe}), but the GCs in the dwarf galaxies (except Fornax 3) mostly have Ba abundances compatible with those in Galactic GCs. As discussed in L12, \citet{Letarte2006} found lower $\mathrm{[Ba/Fe]}$ ratios for three individual stars in Fornax~3 (between $\mathrm{[Ba/Fe]}=+0.09$ and $+0.27$), and although the value found here is lower than that quoted in L12 (due to the modifications in our analysis discussed at the beginning of Sec.~\ref{sec:results}), an offset remains.

The GCs in M33 all have relatively high Ba abundances, $\mathrm{[Ba/Fe]}\approx+0.5$, that are rare among Galactic GCs. It is worth noting that since the Ba lines are quite strong, especially at higher metallicities, they are also relatively sensitive to details of the atmospheres, hyperfine structure, NLTE effects, etc., which might introduce significant systematic uncertainties. In L17 we found that our integrated-light measurements may overestimate the $\mathrm{[Ba/Fe]}$ ratios by 0.20-0.35 dex at the lowest metallicities ($\mathrm{[Fe/H]} \la -2$), compared to literature data for individual stars, although better agreement was found for more metal-rich clusters. From Fig.~\ref{fig:bafe}, the Milky Way GCs and field stars do show a trend of increasing $\mathrm{[Ba/Fe]}$ with metallicity, and if our measurements were shifted downwards by roughly 0.2 dex they would match the Milky Way data quite closely. Thus, the evidence for enhanced $\mathrm{[Ba/Fe]}$ in the M33 GCs should be considered somewhat tentative at this point.

\section{Discussion}  

\subsection{Metallicities}

Our measurements confirm that GCs in dwarf galaxies tend to be relatively metal-poor, compared both with their counterparts in the Milky Way halo and with the field stars in the respective galaxies.  In the Milky Way, 13 out of 152 GCs (slightly less than 9\%) have metallicities $\mathrm{[Fe/H]}<-2$. In Fornax, four out of five GCs fall in this range, and in NGC~147 this is true for four of the five bright GCs for which we have spectroscopic metallicities. In NGC~6822 we only have data for two GCs, but again both of these fall below the peak of the field star metallicity distribution at $\mathrm{[Fe/H]}=-0.84$ \citep{Swan2016}. Nevertheless, it may be notable that none of the GCs for which we have obtained accurate metallicity determinations until now have metallicities below $\mathrm{[Fe/H]}\approx-2.5$.

The difference between the metallicity distributions of the field stars and GCs is striking: at metallicities above $\mathrm{[Fe/H]}=-2$, Fornax has only one GC (Fornax~4), which accounts for about 0.4\% of the luminosity of the field stars. In NGC~147 the fraction is still lower, 0.04\%. These numbers may be compared with the corresponding fractions of $\sim20\%$ and $\sim6\%$ at the low-metallicity end.
It is illustrative to compare these ratios with the specific GC frequencies and -luminosities in larger galaxies. \citet{Forbes2001b} noted that the ``bulge specific frequency'' is about constant at $S_N \approx 1$ for spirals and ellipticals. For an average GC luminosity of $L_V \simeq 10^5 L_{\odot, V}$, this corresponds to a specific GC luminosity of about 0.1\%. This number is not very different from those found for the metal-rich components of Fornax and NGC~147, especially when the poor statistics are taken into account.

\citet{Lamers2017} proposed that differences in GC specific luminosity are caused primarily by more efficient destruction of metal-rich clusters in their gas-rich natal environments. However, they also noted that more continuous, low-level star formation may disfavour the formation of massive clusters. 
In many present-day star forming environments, the mass function of young clusters can be approximated by a Schechter-like function $\mathrm{d} N/\mathrm{d} M \propto M^{-2} \exp (-M/M_c)$ \citep{Larsen2009} where the truncation mass $M_c$ scales with the star formation rate surface density, $\Sigma_\mathrm{SFR}$ \citep{Johnson2017}. Hence, a low $\Sigma_\mathrm{SFR}$ might account for a lack of massive cluster formation. 

The star formation history of the Fornax dwarf has been studied in detail by \citet{DeBoer2012}, who found a peak star formation rate of (3--4)$\times10^{-3}\, M_\odot$ yr$^{-1}$ between 5 and 10 Gyr ago. Combined with a half-number radius of 19\arcmin\ or 750 pc \citep{Battaglia2006}, this gives an average $\Sigma_\mathrm{SFR} \simeq 1\times10^{-3} \, M_\odot \, \mathrm{yr}^{-1} \, \mathrm{kpc}^{-2}$ within the half-number radius. This is about an order of magnitude lower than the $\Sigma_\mathrm{SFR}$ values typical of normal disc galaxies in the Local Universe \citep{Kennicutt1998a}. From the relation in \citet{Johnson2017}, this $\Sigma_\mathrm{SFR}$ would correspond to a truncation mass of $M_c \simeq 4000 \, M_\odot$, which would naturally explain why Fornax did not form any massive clusters for most of its lifetime. 

In NGC~147, about half of the stars appear to have formed during the peak of star formation between 5--7 Gyr ago \citep{Geha2015}. Assuming $M_V=-16.5$, NGC~147 is about 3.3 mag brighter than Fornax, and if we scale the stellar mass of Fornax \citep{Coleman2008} accordingly (ignoring differences in the detailed star formation histories), then the total stellar mass of NGC~147 is $M\sim1.3\times10^9 M_\odot$, which translates to a peak SFR of $\sim 0.32 \, M_\odot \, \mathrm{yr}^{-1}$. For a half-light radius of 1.4 kpc \citep{Crnojevic2014}, we then get $\Sigma_\mathrm{SFR} = 0.025 M_\odot \, \mathrm{yr}^{-1} \, \mathrm{kpc}^{-2}$ within the half-light radius, which is more than an order of magnitude higher than for Fornax. Using the relation in \citet{Johnson2017}, this gives a truncation mass of $M_c = 1.3\times10^5 M_\odot$, suggesting that some relatively massive clusters might have formed in NGC~147 during this epoch. Given that there are few, if any, GCs associated with the metal-rich field stars in NGC~147, cluster disruption may indeed have played a more important role in NGC~147. 

The more important question may be why dwarf galaxies were so proficient at forming metal-poor GCs that managed to survive until today, rather than why they did not form any metal-rich ones. Here, it is worth noting that the relatively crude age resolution even in the detailed CMD-based study by \citet{DeBoer2012} makes it difficult to distinguish between smooth SFHs and more bursty ones. Hence, the metal-poor GCs could have formed during brief, intense bursts of star formation early in the evolutionary histories of the galaxies.

One possible way to distinguish between low formation efficiencies and high disruption efficiencies would be to look for stars with GC-like abundance patterns among metal-rich field stars. If such stars exist, it would favour a picture in which massive GCs did form, but were subsequently destroyed. So far, no significant populations of field stars with GC-like abundance patterns have been found in dwarf galaxies \citep{Lemasle2014,Lardo2016,Suda2017}, but the samples may still be too small to rule out an enriched fraction of a few percent, similar to that seen in the Milky Way halo \citep{Martell2011,Martell2016}.

\subsection{Detailed abundances}

\subsubsection{Overall trends and chemical enrichment histories}

In terms of their detailed abundance patterns, the metal-poor GCs in the dwarf galaxies studied here are similar to metal-poor GCs in the Milky Way halo. The $\alpha$-elements (Ca, Ti) are enhanced by $\sim0.3$ dex relative to Solar-scaled abundance patterns.
While usually classified as an iron-peak element, Sc also tends to be enhanced in metal-poor, $\alpha$-enhanced populations in the Milky Way \citep{Nissen2000,Fishlock2017}. In our observations, we see the same pattern of slightly super-solar $\mathrm{[Sc/Fe]}$ ratios in the metal-poor, $\alpha$-enhanced GCs. This suggests that, at least to first order, the early chemical enrichment histories were quite similar in these different environments.

More significant differences compared to Milky Way GCs start to become apparent for the more metal-rich clusters, Fornax~4 and NGC~6822~SC7. We have already noted the general similarity of the abundance patterns in these clusters to those observed in field stars in nearby dwarf galaxies.
Thus, it appears that GCs are indeed useful tracers of the chemical evolutionary histories in their parent galaxies. The important caveat here, of course, is the difference in the metallicity distributions of field stars and GCs, which means that it is not straight-forward to make general statements about \emph{distributions} of abundances and abundance ratios from GCs. However, GCs can provide useful information on \emph{trends}, such as $[\alpha/\mathrm{Fe}]$ vs.\ $\mathrm{[Fe/H]}$ and correlations between these quantities and \emph{age} (to the extent that ages of the GCs are known). 

\subsubsection{Evidence for multiple populations: Mg spreads and the Na-Ni relation}

   \begin{figure}
   \centering
   \includegraphics[width=\columnwidth]{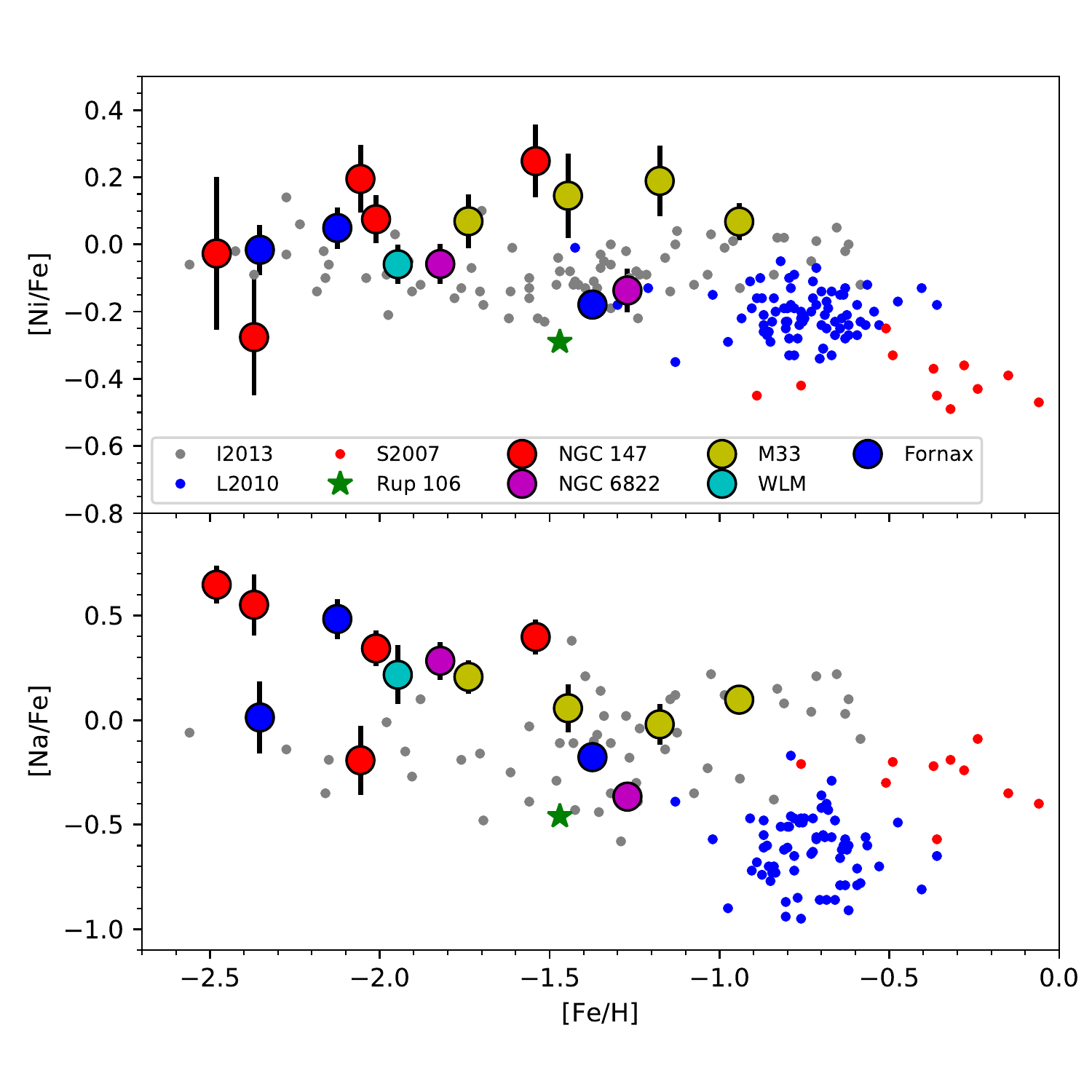}
      \caption{
         \label{fig:nani2}$\mathrm{[Ni/Fe]}$ and $\mathrm{[Na/Fe]}$ with data for stars in the 
         Fornax dSph \citep[blue dots,][]{Letarte2010} and in Sagittarius \citep[red dots,][]{Sbordone2007} also included.
         }
   \end{figure}

   \begin{figure}
   \centering
   \includegraphics[width=\columnwidth]{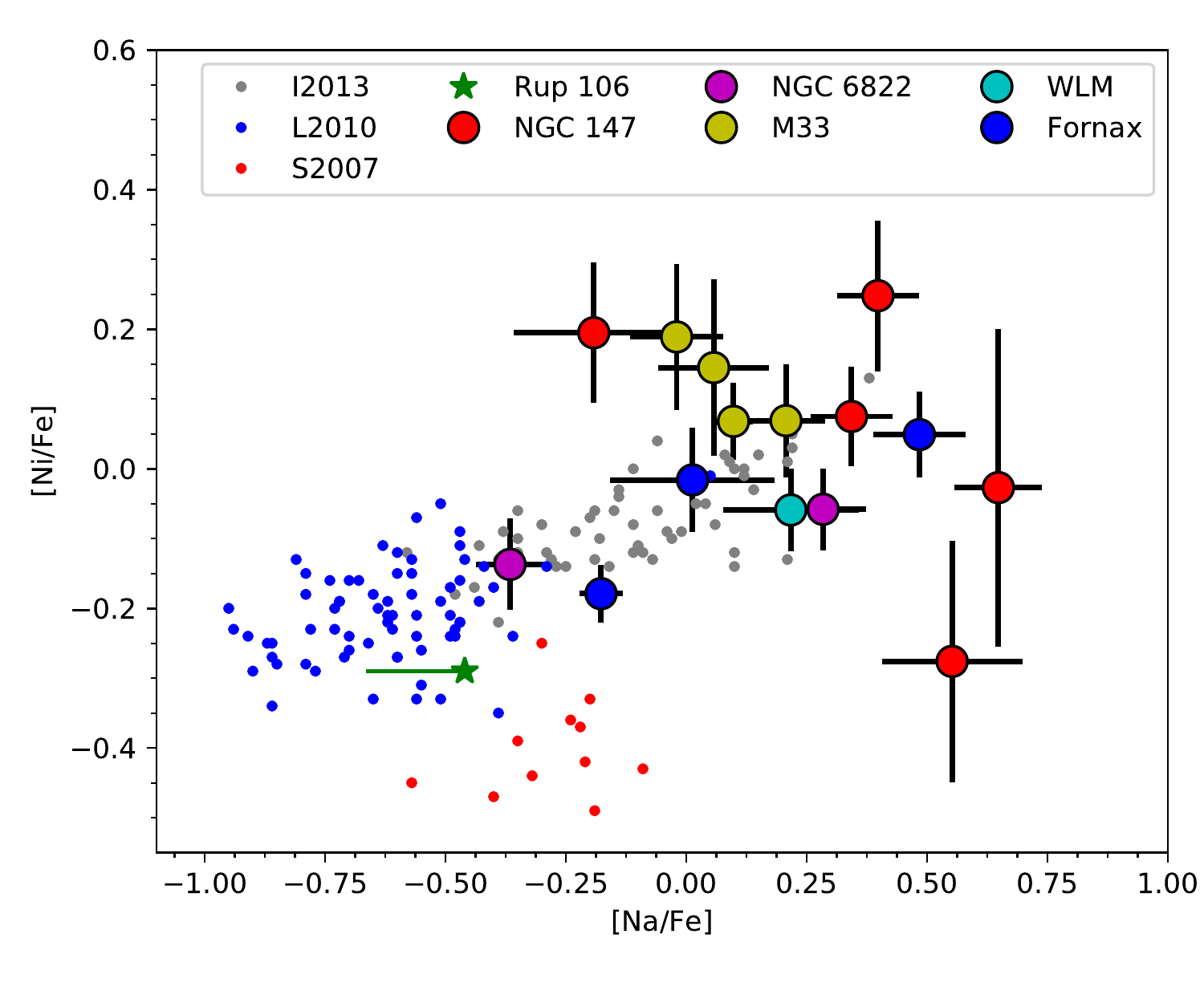}
      \caption{
         \label{fig:nani}$\mathrm{[Ni/Fe]}$ vs.\ $\mathrm{[Na/Fe]}$. Also included: stars in the Milky Way \citep[grey dots,][]{Ishigaki2013}, in the Fornax dSph \citep[blue dots,][]{Letarte2010}, and in Sagittarius \citep[red dots,][]{Sbordone2007}.
         }
   \end{figure}

In most clusters, Mg follows the same abundance patterns as the other $\alpha$-elements. In the few (3--5) cases where significant deviations occur, they are in the sense that Mg is depleted relative to Ca and Ti. This may suggest the presence of Mg abundance anomalies within the affected clusters, although significantly depleted $\mathrm{[Mg/Fe]}$ ratios would require much more extreme Mg anomalies than those observed in most Milky Way GCs. In M15, which has a relatively extended Mg-Al anticorrelation \citep{Sneden1997}, the integrated-light $\mathrm{[Mg/Fe]}$ ratio is only slightly depleted with respect to Ca and Ti, with $[\mathrm{Mg}/\langle\mathrm{Ca,Ti}\rangle] = -0.13$ according to our integrated-light measurements (L17; see also Fig.~\ref{fig:mgafe}). To produce a depletion of more than 0.2 dex in the average $\mathrm{[Mg/Fe]}$, more extreme Mg anomalies would be required, such as those seen in the Galactic GC NGC~2419 for which $\mathrm{[Mg/Fe]}$ ratios as low as $-1$ dex are observed \citep{Mucciarelli2012a,Cohen2011a,Cohen2012}. For this cluster, \citet{Mucciarelli2012a} found 
$\langle \mathrm{[Mg/Fe]} \rangle = +0.05\pm0.08$,
$\langle \mathrm{[Ca/Fe]} \rangle = +0.46\pm0.01$, and
$\langle \mathrm{[Ti/Fe]} \rangle = +0.29\pm0.02$, indicating a Mg depletion of 0.41 and 0.24 dex relative to Ca and Ti, respectively, which is comparable to the depletion seen in the extragalactic GCs.

The clusters in our sample that have depleted $[\mathrm{Mg}/\langle\mathrm{Ca,Ti}\rangle]$ ratios do not appear to have particularly unusual abundances of Na or other elements. Within the Milky Way GC population, some clusters with substantial Mg spreads do display anti-correlated $\mathrm{[Na/Fe]}$ and $\mathrm{[Mg/Fe]}$ abundances \citep[e.g. M15; ][]{Sneden1997}, while others do not \citep[e.g. NGC~2419; ][]{Cohen2012}. Furthermore, while Na spreads are nearly ubiquitous, significant Mg spreads are not \citep{Carretta2009}. 
Perhaps, with larger samples, patterns will emerge, but at this point the origin of large Mg spreads and their exact relation to other abundance ratios remain unclear. 

More generally, the $\mathrm{[Na/Fe]}$ ratios represent a potentially promising diagnostic for identifying multiple populations in extragalactic GCs. The spread in $\mathrm{[Na/Fe]}$ is substantial in most Galactic GCs \citep[$\sim0.5$ dex;][]{Carretta2009}, and the $\mathrm{[Na/Fe]}$ ratios observed in integrated light observations are elevated well above the level seen in field stars \citep[L17;][]{Sakari2013,Colucci2017}. Our measurements here show a similar enhancement in most of the metal-poor GCs, whereas the more metal-rich clusters tend to have lower $\mathrm{[Na/Fe]}$ ratios.
A better understanding of possible systematics in the Na abundance measurements, such as non-LTE effects, would be desirable to quantify the significance of such trends.
A complication here is that the field star Na abundances are not necessarily the same in the extragalactic environments as in the Milky Way halo \citep[Sect.~\ref{sec:abun};][]{Shetrone2003,Cohen2009,Cohen2010,Letarte2010,Lemasle2014,Hendricks2016}. 

For extragalactic applications, it may not always be feasible to obtain detailed abundance measurements for individual field stars, and it would therefore be desirable with other methods to establish the reference Na abundances against which data for GCs could be compared. In this regard, the $\mathrm{[Ni/Fe]}$ ratio may be a useful option, as it is correlated with $\mathrm{[Na/Fe]}$ in Galactic field stars \citep{Nissen1997,Nissen2010}. The correlation has also been extended to dwarf galaxies \citep{Tolstoy2009,Cohen2009,Cohen2010,Lemasle2014}. The origin of the Na-Ni correlation in field stars is suggested to be linked to explosive nucleosynthesis in type II SNe \citep{Letarte2010}, so it should be independent of the hydrogen-burning processes that are usually invoked to explain abundance anomalies in GCs \citep[e.g.][]{Gratton2012}. However, since Ni is also produced in large amounts in type Ia SNe, the exact behaviour of the Na-Ni correlation is expected to depend on the detailed chemical enrichment history, and may well differ from one galaxy to another.

\begin{table}
\caption{Differences between abundances based on the most recent Kurucz line list and those in Table~\ref{tab:abun}.}
\label{tab:kurdiff}
\centering
\begin{tabular}{l c c}
\hline\hline
        & $\langle\Delta_\mathrm{Kur-CH04}\rangle$ & $\sigma$ \\\hline
$\mathrm{[Fe/H]}$  & $-0.016$ & 0.014 \\
$\mathrm{[Na/Fe]}$ & $+0.017$ & 0.014 \\
$\mathrm{[Mg/Fe]}$ & $+0.048$ & 0.025 \\
$\mathrm{[Ca/Fe]}$ & $+0.023$ & 0.016 \\
$\mathrm{[Sc/Fe]}$ & $+0.099$ & 0.049 \\
$\mathrm{[Ti/Fe]}$ & $-0.017$ & 0.036 \\
$\mathrm{[Cr/Fe]}$ & $+0.035$ & 0.037 \\
$\mathrm{[Mn/Fe]}$ & $+0.069$ & 0.057 \\
$\mathrm{[Ni/Fe]}$ & $-0.054$ & 0.157 \\
$\mathrm{[Ba/Fe]}$ & $-0.055$ & 0.074 \\
\hline
\end{tabular}
\end{table}

Fig.~\ref{fig:nani2} shows our measurements of $\mathrm{[Na/Fe]}$ and $\mathrm{[Ni/Fe]}$ vs. $\mathrm{[Fe/H]}$, now together with data for field stars in the Fornax dwarf \citep{Letarte2010} and Sagittarius \citep{Sbordone2007}, and in Fig.~\ref{fig:nani} we plot the same data in the $\mathrm{[Na/Fe]}$ vs. $\mathrm{[Ni/Fe]}$ plane. At this stage, it is not clear that unambiguous conclusions about Na enrichment in the GCs can be drawn directly from Fig.~\ref{fig:nani}.  Clusters with additional Na enrichment (beyond that following from the Na-Ni correlation) should lie to the right of the sequence formed by the field stars, but there is no clear tendency in Fig.~\ref{fig:nani} for this to actually be the case. Instead, there is a large scatter around the locus occupied by the stars. Some of this scatter may well be due to observational uncertainties - in particular, for the two most metal-poor GCs in NGC~147 (N147-PA-1 and Hodge~III), most of the individual measurements of $\mathrm{[Ni/Fe]}$ are upper limits or did not converge, and the remaining 4--5 measurements for each of these clusters show a large dispersion. We have noted (Sec.~\ref{sec:nafe}) that the Na abundances may be underestimated by $\sim0.1$ dex, and an increase of the $\mathrm{[Na/Fe]}$ ratios by this amount would indeed shift the data points in Fig.~\ref{fig:nani} to the right. Alternatively, it is also possible that the $\mathrm{[Ni/Fe]}$ ratios are systematically overestimated, and we note that the line-to-line scatter on the Ni abundances in Table~\ref{tab:abun} is indeed quite large for most clusters.

A complicating factor in this comparison is that the different galaxies occupy distinct loci in Fig.~\ref{fig:nani}, with the \citet{Sbordone2007} measurements for Sagittarius being offset from the others.  While this might be due to different enrichment histories, we note that \citet{Hasselquist2017} find Ni and Na abundance ratios for Sagittarius closer to those seen in Fornax. They find a mean $\mathrm{[Na/Fe]}\simeq-0.6$ (with a large scatter) and $\mathrm{[Ni/Fe]}\simeq-0.2$ for $-0.8 < \mathrm{[Fe/H]} < 0$, which would largely remove the offset between the Fornax and Sagittarius measurements. 
Another problem in making these comparisons is that the field stars in Fornax and Sagittarius, for which Na and Ni abundances are available, generally have higher metallicities than the GCs in our sample (Fig.~\ref{fig:nani2}), so it is unclear whether the more metal-poor stars in the dwarf galaxies follow the same $\mathrm{[Ni/Fe]}$ vs.\ $\mathrm{[Na/Fe]}$ relation as stars in the Milky Way.

\subsubsection{Analysis with Kurucz line list}

   \begin{figure}
   \centering
   \includegraphics[width=\columnwidth]{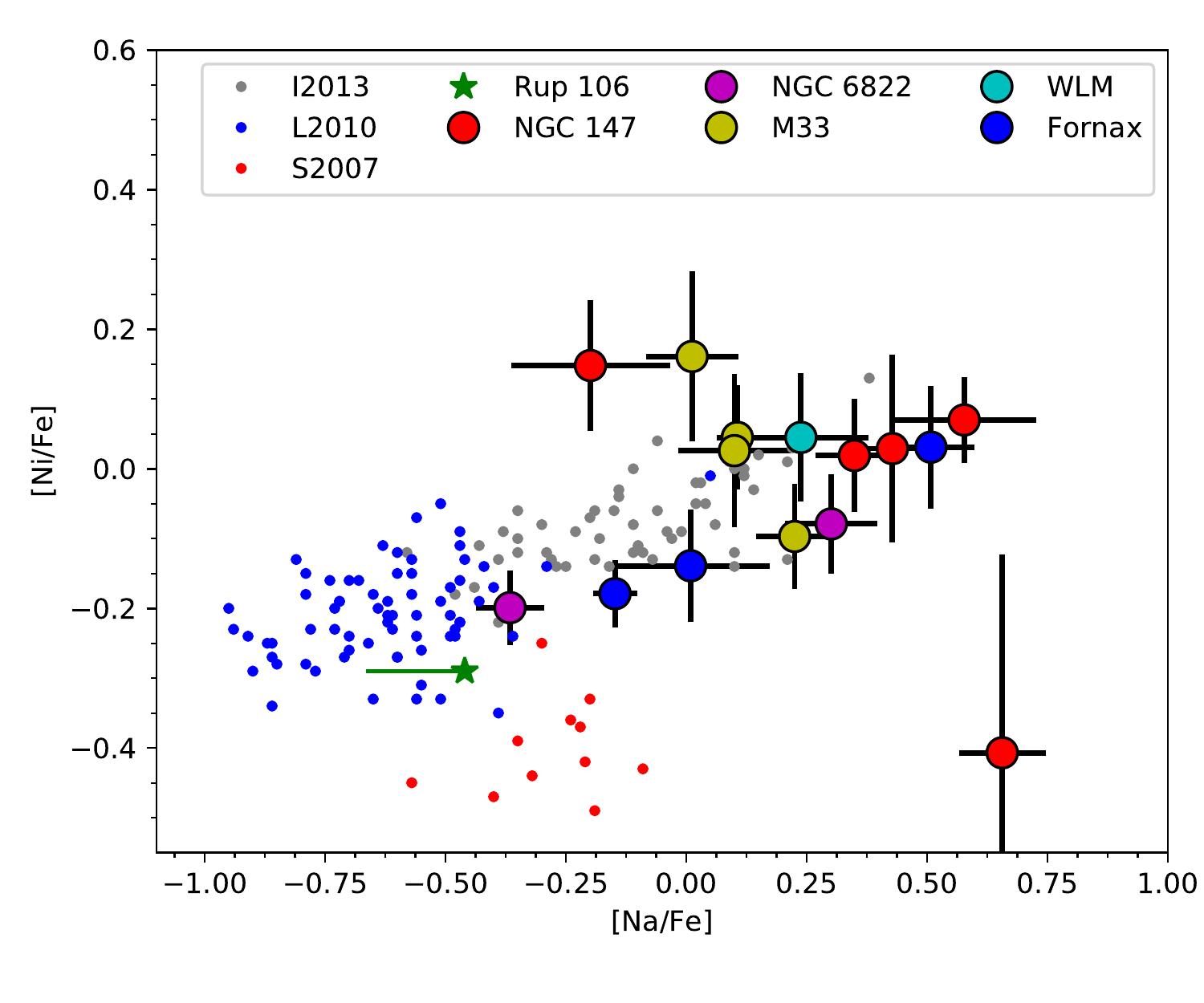}
      \caption{
         \label{fig:nani_kur}Same as Fig.~\ref{fig:nani}, but using the most recent line list from the Kurucz web site.
         }
   \end{figure}

An important source of systematic uncertainty is the input line list used in the analysis. As recently pointed out by \citet{Laverick2017}, the atomic data included in different databases can differ substantially. Differences in the input line list can affect our abundances not only through the data for the measured individual lines themselves, but also via other nearby lines that may be blended with a particular feature of interest or affect the scaling of the spectra when carrying out the fits.

In L17 we compared results based on our ``standard'' line list with fits based on a more recent version of the Kurucz list. 
The Kurucz list is updated regularly and here we have carried out an additional set of fits to the GC spectra in the current sample using the most recent version (dated 08-Oct-2017). The results of these fits are given in the Appendix (Table~\ref{tab:abunkur}) and Table~\ref{tab:kurdiff} lists the mean differences between the abundances in Table~\ref{tab:abun} and those in Table~\ref{tab:abunkur}, as well as the cluster-to-cluster dispersions.
As found in L17, the differences are fairly small, in all cases less than 0.1 dex, and the conclusions based on our modified version of the CH04 line list remain mostly unchanged. The $\mathrm{[Na/Fe]}$ ratios increase very slightly, by 0.017 dex (mostly driven by a corresponding decrease in $\mathrm{[Fe/H]}$), and the average $\mathrm{[Mg/Fe]}$ ratios increase by about 0.05~dex, but the three Mg-depleted clusters remain so also when using the new Kurucz line list. The $\mathrm{[Ba/Fe]}$ ratios decrease by about 0.05~dex, but this is not enough to remove the offset in Fig.~\ref{fig:bafe}. Fornax~4 and NGC~6822-SC7 remain less $\alpha$-enhanced than their counterparts at similar metallicities in M33 and the Milky Way.

The scatter in Table~\ref{tab:kurdiff} is again particularly large for $\mathrm{[Ni/Fe]}$, and it is thus worth examining how this may affect Fig.~\ref{fig:nani}. In Fig.~\ref{fig:nani_kur} we show the $\mathrm{[Ni/Fe]}$ vs. $\mathrm{[Na/Fe]}$ plot once again, but using the measurements based on the Kurucz line list. There is now a somewhat clearer tendency for the GC measurements to fall to the right of, or below, the relation followed by the stars, although a substantial scatter remains and there are still a couple of clusters on the ``wrong'' side of the relation. Note that we still haven't accounted for possible systematic errors on $\mathrm{[Na/Fe]}$, which might shift the data points further towards the right.

\subsubsection{NGC~6822~SC7: A ``twin'' of Ruprecht 106}

Finally, we comment on the remarkable similarity of the abundance patterns of NGC~6822~SC7 to those in the peculiar Galactic GC Ruprecht~106, for which we have indicated the abundance measurements by \citet{Villanova2013} with green asterisks in Fig.~\ref{fig:alphafe} and Figs.~\ref{fig:nafe}-\ref{fig:bafe}. Like NGC~6822~SC7, Ruprecht~106 shows no enhancement of the $\alpha$-elements, and Na, Sc, and Ni are all significantly depleted compared to more typical Milky Way GC abundance patterns. The one dissimilarity is for Ba, which is depleted in Ruprecht~106 but about normal in NGC~6822~SC7. The suggestion that the unusual abundance patterns in Ruprecht~106 indicate an extragalactic origin \citep{Villanova2013} may be supported by the similarity to NGC~6822~SC7. 

A noteworthy property of Ruprecht~106 is that it appears to be chemically \emph{homogeneous}, i.e., it shows no evidence for multiple stellar populations. The similarity of the abundance patterns, in particular Na and Ni, might suggest that this is true also for NGC~6822~SC7. 
Again, Fig.~\ref{fig:nani} does not offer a clear-cut conclusion: in fact, Ruprecht~106 is offset with respect to the mean $\mathrm{[Ni/Fe]}$ vs.\ $\mathrm{[Na/Fe]}$ relation for field stars in Fornax in the direction that would suggest Na-enrichment. (We have used the LTE Na abundance from \citet{Villanova2013} for comparison with our measurements; using their NLTE abundance would shift the point as indicated by the green line, closer to the field stars.) However, if anything the evidence against multiple populations would then appear \emph{stronger} in NGC~6822~SC7, which falls closer to the locus of the field stars. \citet{Villanova2013} suggested that the apparent absence of multiple populations in Rup~106 might be related to its relatively low mass. For NGC~6822~SC7, such an explanation cannot be invoked. The velocity dispersion is the highest among the clusters studied here, and the absolute magnitude is $M_V=-8.6$, corresponding to a mass of $\sim3.5\times10^5\, M_\odot$ (assuming a mass-to-light ratio of $\sim1.5$). However, we emphasise that the current evidence does not allow us to establish with certainty that multiple populations are absent from SC7.

\section{Summary}

We have presented new integrated-light measurements of chemical abundances for 11 globular clusters in NGC~147, NGC~6822, and Messier 33. Combined with our previous data for GCs in the Fornax and WLM galaxies, we have now carried out such measurements for 15 Local Group GCs. Our principal findings are as follows:

\begin{itemize}
\item The GCs in the dwarf galaxies tend to be more metal-poor on average than GCs in the Galactic halo. Nevertheless, it may be significant that no GCs with $\mathrm{[Fe/H]}\la-2.5$ have yet been found, neither in the Milky Way nor among the clusters we have observed so far.
\item 
The $[\alpha/\mathrm{Fe}]$ ratios behave differently as a function of metallicity in the dwarf galaxies and M33: While all metal-poor clusters ($\mathrm{[Fe/H]}\lesssim-1.5$) are $\alpha$-enhanced at about the same level as Milky Way GCs, the more metal-rich GCs in the dwarf galaxies have close to Solar $[\alpha/\mathrm{Fe}]$ ratios, while those in M33 remain $\alpha$-enhanced.  
\item
In most clusters, Mg follows the other $\alpha$-elements, Ca and Ti. A small subset of the clusters have depleted $\mathrm{[Mg/Fe]}$ ratios (by 0.2--0.3 dex) relative to Ca and Ti, as found previously in other extragalactic GCs. 
\item
In the metal-poor GCs, $\mathrm{[Na/Fe]}$ is elevated as in Milky Way GCs, consistent with the presence of Na-rich stars in the clusters. In the more metal-rich clusters, the $\mathrm{[Na/Fe]}$ ratios are generally lower than in Milky Way GCs, in particular in Fornax~4 and the cluster SC7 in NGC~6822. However, since we lack information about Na abundances at the same metallicities in the field stars, the implications for the presence of multiple populations are unclear.
\item
Fornax~4 and SC7 also have low $\mathrm{[Ni/Fe]}$ ratios, as observed also in field stars in nearby dwarf galaxies.
\item
The $\mathrm{[Ba/Fe]}$ ratios show a large spread, but may be systematically elevated in the M33 GCs. Interpretation of such an offset is complicated by the fact that Ba can be produced by both the $r$- and $s$-process. Constraining the relative important of the $r$- and $s$-process requires information about other $n$-capture elements, such as Eu. This, however, requires better sensitivity in the blue part of the spectrum than we have achieved here.
\end{itemize}

In the dwarf galaxies, the trends followed by the GCs are generally consistent with data for field stars in nearby dwarf galaxies, and our measurements provide the first evidence of the ``knee'' in the $[\alpha/\mathrm{Fe}]$ vs. $\mathrm{[Fe/H]}$ relation in NGC~6822.  The measurements presented here also provide the first detailed information on chemical abundances in the M33 halo. The $\alpha$-elements in the M33 GCs follow patterns similar to those seen in Milky Way GCs, which suggests that the M33 halo underwent relatively rapid chemical enrichment, dominated by Type II SN nucleosynthesis. 
The tentative nature of the results regarding barium must be stressed, but we note that elevated $\mathrm{[Ba/Fe]}$ ratios would be reminiscent of those observed in some dwarf galaxies, possibly indicating a stronger contribution to chemical enrichment by AGB stars in M33 than in the Milky Way halo. 
The combination of Na and Ni abundance measurements may ultimately turn out to be useful for detecting multiple populations in integrated-light observations, provided that an accuracy of $\sim0.1$~dex or better can be reached for the integrated-light abundance measurements.
However, this also requires a better understanding of the Na-Ni relation in extragalactic environments.

\begin{acknowledgements}
We thank the anonymous referee for a careful reading of the manuscript and a number of helpful comments.
Some of the data presented herein were obtained at the W. M. Keck Observatory, which is operated as a scientific partnership among the California Institute of Technology, the University of California and the National Aeronautics and Space Administration. The Observatory was made possible by the generous financial support of the W. M. Keck Foundation.
This research has made use of the NASA/IPAC Extragalactic Database (NED) which is operated by the Jet Propulsion Laboratory, California Institute of Technology, under contract with the National Aeronautics and Space Administration.
This work has made use of the VALD database, operated at Uppsala University, the Institute of Astronomy RAS in Moscow, and the University of Vienna. This research has made use of NASA's Astrophysics Data System Bibliographic Services.
JB acknowledges support by NSF grant AST-1518294 and HST grant GO-13295.001-A.
JS acknowledges support by NSF grant AST-1514763 and a Packard Fellowship.
\end{acknowledgements}

\bibliographystyle{aa}
\bibliography{refs.bib}

\begin{thebibliography}{137}
\expandafter\ifx\csname natexlab\endcsname\relax\def\natexlab#1{#1}\fi

\bibitem[{Battaglia {et~al.}(2006)Battaglia, Tolstoy, Helmi, Irwin, Letarte,
  Jablonka, Hill, Venn, Shetrone, Arimoto, Primas, Kaufer, Francois, Szeifert,
  Abel, \& Sadakane}]{Battaglia2006}
Battaglia, G., Tolstoy, E., Helmi, A., {et~al.} 2006, A\&A, 459, 423

\bibitem[{Battistini \& Bensby(2015)}]{Battistini2015}
Battistini, C. \& Bensby, T. 2015, A\&A, 577, A9

\bibitem[{Beasley {et~al.}(2015)Beasley, {San Roman}, Gallart, Sarajedini, \&
  Aparicio}]{Beasley2015}
Beasley, M.~A., {San Roman}, I., Gallart, C., Sarajedini, A., \& Aparicio, A.
  2015, MNRAS, 451, 3400

\bibitem[{Bensby {et~al.}(2003)Bensby, Feltzing, \&
  Lundstr{\"{o}}m}]{Bensby2003}
Bensby, T., Feltzing, S., \& Lundstr{\"{o}}m, I. 2003, A\&A, 410, 527

\bibitem[{Bensby {et~al.}(2014)Bensby, Feltzing, \& Oey}]{Bensby2014}
Bensby, T., Feltzing, S., \& Oey, M.~S. 2014, A\&A, 562, A71

\bibitem[{Bergemann \& Gehren(2008)}]{Bergemann2008}
Bergemann, M. \& Gehren, T. 2008, A\&A, 492, 823

\bibitem[{Bressan {et~al.}(2012)Bressan, Marigo, Girardi, Salasnich, Cero,
  Rubele, \& Nanni}]{Bressan2012}
Bressan, A., Marigo, P., Girardi, L., {et~al.} 2012, MNRAS, 427, 127

\bibitem[{Brodie \& Huchra(1991)}]{Brodie1991}
Brodie, J.~P. \& Huchra, J.~P. 1991, ApJ, 379, 157

\bibitem[{Brooks {et~al.}(2004)Brooks, Wilson, \& Harris}]{Brooks2004}
Brooks, R.~S., Wilson, C.~D., \& Harris, W.~E. 2004, AJ, 128, 237

\bibitem[{Cabrera-Ziri {et~al.}(2016)Cabrera-Ziri, Lardo, Davies, Bastian,
  Beccari, Larsen, \& Hernandez}]{Cabrera-Ziri2016}
Cabrera-Ziri, I., Lardo, C., Davies, B., {et~al.} 2016, MNRAS, 460, 1869

\bibitem[{Carretta {et~al.}(2009)Carretta, Bragaglia, Gratton, \&
  Lucatello}]{Carretta2009}
Carretta, E., Bragaglia, A., Gratton, R., \& Lucatello, S. 2009, A\&A, 505, 139

\bibitem[{Castelli \& Hubrig(2004)}]{Castelli2004}
Castelli, F. \& Hubrig, S. 2004, A\&A, 425, 263

\bibitem[{Chandar {et~al.}(2002)Chandar, Bianchi, Ford, \&
  Sarajedini}]{Chandar2002}
Chandar, R., Bianchi, L., Ford, H.~C., \& Sarajedini, A. 2002, ApJ, 564, 712

\bibitem[{Chiappini {et~al.}(2001)Chiappini, Matteucci, \&
  Romano}]{Chiappini2001}
Chiappini, C., Matteucci, F., \& Romano, D. 2001, ApJ, 554, 1044

\bibitem[{Christian \& Schommer(1982)}]{Christian1982}
Christian, C.~A. \& Schommer, R.~A. 1982, ApJS, 49, 405

\bibitem[{Cohen {et~al.}(2011)Cohen, Huang, \& Kirby}]{Cohen2011a}
Cohen, J., Huang, W., \& Kirby, E. 2011, ApJ, 740, 60

\bibitem[{Cohen(1978)}]{Cohen1978}
Cohen, J.~G. 1978, ApJ, 223, 487

\bibitem[{Cohen \& Huang(2009)}]{Cohen2009}
Cohen, J.~G. \& Huang, W. 2009, ApJ, 701, 1053

\bibitem[{Cohen \& Huang(2010)}]{Cohen2010}
Cohen, J.~G. \& Huang, W. 2010, ApJ, 719, 931

\bibitem[{Cohen \& Kirby(2012)}]{Cohen2012}
Cohen, J.~G. \& Kirby, E.~N. 2012, ApJ, 760, 86

\bibitem[{Coleman \& de~Jong(2008)}]{Coleman2008}
Coleman, M.~G. \& de~Jong, J. T.~A. 2008, ApJ, 685, 933

\bibitem[{Colucci \& Bernstein(2011)}]{Colucci2011}
Colucci, J. \& Bernstein, R. 2011, EAS Publications Series, 48, 275

\bibitem[{Colucci {et~al.}(2009)Colucci, Bernstein, Cameron, McWilliam, \&
  Cohen}]{Colucci2009}
Colucci, J.~E., Bernstein, R.~A., Cameron, S., McWilliam, A., \& Cohen, J.~G.
  2009, ApJ, 704, 385

\bibitem[{Colucci {et~al.}(2014)Colucci, Bernstein, \& Cohen}]{Colucci2014}
Colucci, J.~E., Bernstein, R.~A., \& Cohen, J.~G. 2014, ApJ, 797, 116

\bibitem[{Colucci {et~al.}(2017)Colucci, Bernstein, \& McWilliam}]{Colucci2017}
Colucci, J.~E., Bernstein, R.~A., \& McWilliam, A. 2017, ApJ, 834, 105

\bibitem[{Conroy {et~al.}(2014)Conroy, Graves, \& van Dokkum}]{Conroy2014}
Conroy, C., Graves, G.~J., \& van Dokkum, P.~G. 2014, ApJ, 780, 33

\bibitem[{Crnojevi{\'{c}} {et~al.}(2014)Crnojevi{\'{c}}, Ferguson, Irwin,
  McConnachie, Bernard, Fardal, Ibata, Lewis, Martin, Navarro, No{\"{e}}l, \&
  Pasetto}]{Crnojevic2014}
Crnojevi{\'{c}}, D., Ferguson, A. M.~N., Irwin, M.~J., {et~al.} 2014, MNRAS,
  445, 3862

\bibitem[{{Da Costa} \& Mould(1988)}]{DaCosta1988}
{Da Costa}, G.~S. \& Mould, J.~R. 1988, ApJ, 334, 159

\bibitem[{de~Boer {et~al.}(2012)de~Boer, Tolstoy, Hill, Saha, Olszewski, Mateo,
  Starkenburg, Battaglia, \& Walker}]{DeBoer2012}
de~Boer, T. J.~L., Tolstoy, E., Hill, V., {et~al.} 2012, A\&A, 544, A73

\bibitem[{Dotter {et~al.}(2007)Dotter, Chaboyer, Jevremovi{\'{c}}, Baron,
  Ferguson, Sarajedini, \& Anderson}]{Dotter2007}
Dotter, A., Chaboyer, B., Jevremovi{\'{c}}, D., {et~al.} 2007, AJ, 134, 376

\bibitem[{Edvardsson {et~al.}(1993)Edvardsson, Andersen, Gustafsson, Lambert,
  Nissen, \& Tomkin}]{Edvardsson1993}
Edvardsson, B., Andersen, J., Gustafsson, B., {et~al.} 1993, A\&A, 275, 101

\bibitem[{Feast {et~al.}(2012)Feast, Whitelock, Menzies, \&
  Matsunaga}]{Feast2012}
Feast, M.~W., Whitelock, P.~A., Menzies, J.~W., \& Matsunaga, N. 2012, MNRAS,
  421, 2998

\bibitem[{Fishlock {et~al.}(2017)Fishlock, Yong, Karakas, Alves-Brito,
  Melendez, Nissen, Kobayashi, \& Casey}]{Fishlock2017}
Fishlock, C.~K., Yong, D., Karakas, A.~I., {et~al.} 2017, MNRAS, 466, 4672

\bibitem[{Forbes {et~al.}(2001)Forbes, Brodie, \& Larsen}]{Forbes2001b}
Forbes, D.~A., Brodie, J.~P., \& Larsen, S.~S. 2001, ApJ, 556, L83

\bibitem[{Forbes \& Forte(2001)}]{Forbes2001}
Forbes, D.~A. \& Forte, J.~C. 2001, MNRAS, 322, 257

\bibitem[{Forte {et~al.}(1981)Forte, Strom, \& Strom}]{Forte1981}
Forte, J.~C., Strom, S.~E., \& Strom, K.~M. 1981, ApJ, 245, L9

\bibitem[{Freeman \& Bland-Hawthorn(2002)}]{Freeman2002}
Freeman, K. \& Bland-Hawthorn, J. 2002, ARA\&A, 40, 487

\bibitem[{Fulbright(2000)}]{Fulbright2000}
Fulbright, J.~P. 2000, AJ, 120, 1841

\bibitem[{Gazak {et~al.}(2014)Gazak, Davies, Bastian, Kudritzki, Bergemann,
  Plez, Evans, Patrick, Bresolin, \& Schinnerer}]{Gazak2014}
Gazak, J.~Z., Davies, B., Bastian, N., {et~al.} 2014, ApJ, 787, 142

\bibitem[{Geha {et~al.}(2010)Geha, van~der Marel, Guhathakurta, Gilbert,
  Kalirai, \& Kirby}]{Geha2010}
Geha, M., van~der Marel, R.~P., Guhathakurta, P., {et~al.} 2010, ApJ, 711, 361

\bibitem[{Geha {et~al.}(2015)Geha, Weisz, Grocholski, Dolphin, van~der Marel,
  \& Guhathakurta}]{Geha2015}
Geha, M., Weisz, D., Grocholski, A., {et~al.} 2015, ApJ, 811, 114

\bibitem[{Gratton {et~al.}(2012)Gratton, Carretta, \& Bragaglia}]{Gratton2012}
Gratton, R.~G., Carretta, E., \& Bragaglia, A. 2012, A\&AR, 20, 1

\bibitem[{Gratton {et~al.}(1999)Gratton, Carretta, Eriksson, \&
  Gustafsson}]{Gratton1999}
Gratton, R.~G., Carretta, E., Eriksson, K., \& Gustafsson, B. 1999, A\&A, 350,
  955

\bibitem[{Harris(1996)}]{Harris1996}
Harris, W.~E. 1996, AJ, 112, 1487

\bibitem[{Harris \& Harris(2002)}]{Harris2002}
Harris, W.~E. \& Harris, G. L.~H. 2002, AJ, 123, 3108

\bibitem[{Harris {et~al.}(2013)Harris, Harris, \& Alessi}]{Harris2013}
Harris, W.~E., Harris, G. L.~H., \& Alessi, M. 2013, ApJ, 772, 82

\bibitem[{Harris {et~al.}(2007)Harris, Harris, Layden, \& Wehner}]{Harris2007}
Harris, W.~E., Harris, G. L.~H., Layden, A.~C., \& Wehner, E. M.~H. 2007, ApJ,
  666, 903

\bibitem[{Hasselquist {et~al.}(2017)Hasselquist, Shetrone, Smith, Holtzman,
  McWilliam, Fern{\'{a}}ndez-Trincado, Beers, Majewski, Nidever, Tang, Tissera,
  Alvar, Prieto, Almeida, Anguiano, Battaglia, Carigi, Inglada, Frinchaboy,
  Garc{\'{i}}a-Hern{\'{a}}ndez, Geisler, Minniti, Placco, Schultheis, Sobeck,
  \& Villanova}]{Hasselquist2017}
Hasselquist, S., Shetrone, M., Smith, V., {et~al.} 2017, ApJ, 845, 162

\bibitem[{Helmi(2008)}]{Helmi2008}
Helmi, A. 2008, A\&AR, 15, 145

\bibitem[{Hendricks {et~al.}(2016)Hendricks, Boeche, Johnson, Frank, Koch,
  Mateo, \& Bailey}]{Hendricks2016}
Hendricks, B., Boeche, C., Johnson, C.~I., {et~al.} 2016, A\&A, 585, A86

\bibitem[{Hernandez {et~al.}(2017)Hernandez, Larsen, Trager, Groot, \&
  Kaper}]{Hernandez2017}
Hernandez, S., Larsen, S., Trager, S., Groot, P., \& Kaper, L. 2017, A\&A, 603,
  A119

\bibitem[{Hernandez {et~al.}(2018)Hernandez, Larsen, Trager, Kaper, \&
  Groot}]{Hernandez2017a}
Hernandez, S., Larsen, S., Trager, S., Kaper, L., \& Groot, P. 2018, MNRAS,
  473, 826

\bibitem[{Ho {et~al.}(2014)Ho, Geha, Tollerud, Zinn, Guhathakurta, \&
  Vargas}]{Ho2014}
Ho, N., Geha, M., Tollerud, E.~J., {et~al.} 2014, ApJ, 798, 77

\bibitem[{Homma {et~al.}(2015)Homma, Murayama, Kobayashi, \&
  Taniguchi}]{Homma2015}
Homma, H., Murayama, T., Kobayashi, M. A.~R., \& Taniguchi, Y. 2015, ApJ, 799,
  230

\bibitem[{Hoyle(1954)}]{Hoyle1954}
Hoyle, F. 1954, ApJS, 1, 121

\bibitem[{Hubble(1925)}]{Hubble1925}
Hubble, E.~P. 1925, ApJ, 62, 409

\bibitem[{Huxor {et~al.}(2013)Huxor, Ferguson, Veljanoski, Mackey, \&
  Tanvir}]{Huxor2013}
Huxor, A., Ferguson, A., Veljanoski, J., Mackey, D., \& Tanvir, N. 2013, MNRAS,
  429, 1039

\bibitem[{Hwang {et~al.}(2011)Hwang, Lee, Lee, Park, Park, Kim, \&
  Park}]{Hwang2011}
Hwang, N., Lee, M.~G., Lee, J.~C., {et~al.} 2011, ApJ, 738, 58

\bibitem[{Ishigaki {et~al.}(2013)Ishigaki, Aoki, \& Chiba}]{Ishigaki2013}
Ishigaki, M.~N., Aoki, W., \& Chiba, M. 2013, ApJ, 771, 67

\bibitem[{Johnson {et~al.}(2017)Johnson, Seth, Dalcanton, Beerman, Fouesneau,
  Weisz, Bell, Dolphin, Sandstrom, \& Williams}]{Johnson2017}
Johnson, L.~C., Seth, A.~C., Dalcanton, J.~J., {et~al.} 2017, ApJ, 839, 14

\bibitem[{Kennicutt(1998)}]{Kennicutt1998a}
Kennicutt, J. 1998, ApJ, 498, 541

\bibitem[{Kissler-Patig {et~al.}(2002)Kissler-Patig, Brodie, \&
  Minniti}]{KisslerPatig2002}
Kissler-Patig, M., Brodie, J.~P., \& Minniti, D. 2002, A\&A, 391, 441

\bibitem[{Kobayashi \& Nakasato(2011)}]{Kobayashi2011}
Kobayashi, C. \& Nakasato, N. 2011, ApJ, 729, 16

\bibitem[{Kobayashi {et~al.}(2006)Kobayashi, Umeda, Nomoto, Tominaga, \&
  Ohkubo}]{Kobayashi2006}
Kobayashi, C., Umeda, H., Nomoto, K., Tominaga, N., \& Ohkubo, T. 2006, ApJ,
  653, 1145

\bibitem[{Kramida {et~al.}(2013)Kramida, Ralchenko, \& Reader}]{NIST}
Kramida, A., Ralchenko, Y., \& Reader, J. 2013, {NIST Atomic Spectra Database}

\bibitem[{Kuntschner(2000)}]{Kuntschner2000}
Kuntschner, H. 2000, MNRAS, 315, 184

\bibitem[{Lamers {et~al.}(2017)Lamers, Kruijssen, Bastian, Rejkuba, Hilker, \&
  Kissler-Patig}]{Lamers2017}
Lamers, H. J. G. L.~M., Kruijssen, J. M.~D., Bastian, N., {et~al.} 2017, A\&A,
  606, A85

\bibitem[{Lanfranchi {et~al.}(2006)Lanfranchi, Matteucci, \&
  Cescutti}]{Lanfranchi2006}
Lanfranchi, G.~A., Matteucci, F., \& Cescutti, G. 2006, A\&A, 453, 67

\bibitem[{Lardo {et~al.}(2016)Lardo, Battaglia, Pancino, Romano, de~Boer,
  Starkenburg, Tolstoy, Irwin, Jablonka, \& Tosi}]{Lardo2016}
Lardo, C., Battaglia, G., Pancino, E., {et~al.} 2016, A\&A, 585, A70

\bibitem[{Lardo {et~al.}(2015)Lardo, Davies, Kudritzki, Gazak, Evans, Patrick,
  Bergemann, \& Plez}]{Lardo2015b}
Lardo, C., Davies, B., Kudritzki, R.-P., {et~al.} 2015, ApJ, 812, 160

\bibitem[{Larsen(2009)}]{Larsen2009}
Larsen, S.~S. 2009, A\&A, 494, 539

\bibitem[{Larsen {et~al.}(2014{\natexlab{a}})Larsen, Brodie, Forbes, \&
  Strader}]{Larsen2014}
Larsen, S.~S., Brodie, J.~P., Forbes, D.~A., \& Strader, J. 2014{\natexlab{a}},
  A\&A, 565, A98

\bibitem[{Larsen {et~al.}(2014{\natexlab{b}})Larsen, Brodie, Grundahl, \&
  Strader}]{Larsen2014a}
Larsen, S.~S., Brodie, J.~P., Grundahl, F., \& Strader, J. 2014{\natexlab{b}},
  ApJ, 797, 15

\bibitem[{Larsen {et~al.}(2002)Larsen, Brodie, Sarajedini, \&
  Huchra}]{Larsen2002b}
Larsen, S.~S., Brodie, J.~P., Sarajedini, A., \& Huchra, J.~P. 2002, AJ, 124,
  2615

\bibitem[{Larsen {et~al.}(2012{\natexlab{a}})Larsen, Brodie, \&
  Strader}]{Larsen2012a}
Larsen, S.~S., Brodie, J.~P., \& Strader, J. 2012{\natexlab{a}}, A\&A, 546, A53

\bibitem[{Larsen {et~al.}(2017)Larsen, Brodie, \& Strader}]{Larsen2017}
Larsen, S.~S., Brodie, J.~P., \& Strader, J. 2017, A\&A, 601, A96

\bibitem[{Larsen {et~al.}(2008)Larsen, Origlia, Brodie, \&
  Gallagher}]{Larsen2008a}
Larsen, S.~S., Origlia, L., Brodie, J., \& Gallagher, J.~S. 2008, MNRAS, 383,
  263

\bibitem[{Larsen {et~al.}(2006)Larsen, Origlia, Brodie, \&
  Gallagher}]{Larsen2006b}
Larsen, S.~S., Origlia, L., Brodie, J.~P., \& Gallagher, J.~S. 2006, MNRAS,
  368, L10

\bibitem[{Larsen {et~al.}(2012{\natexlab{b}})Larsen, Strader, \&
  Brodie}]{Larsen2012}
Larsen, S.~S., Strader, J., \& Brodie, J.~P. 2012{\natexlab{b}}, A\&A, 544, L14

\bibitem[{Laverick {et~al.}(2017)Laverick, Lobel, Merle, Royer, Martayan,
  David, Hensberge, \& Thienpont}]{Laverick2017}
Laverick, M., Lobel, A., Merle, T., {et~al.} 2017, A\&A, arXiv prep, 1712.07678

\bibitem[{Lee {et~al.}(2006)Lee, Skillman, \& Venn}]{Lee2006a}
Lee, H., Skillman, E.~D., \& Venn, K.~A. 2006, ApJ, 642, 813

\bibitem[{Lemasle {et~al.}(2014)Lemasle, de~Boer, Hill, Tolstoy, Irwin,
  Jablonka, Venn, Battaglia, Starkenburg, Shetrone, Letarte, Francois, Helmi,
  Primas, Kaufer, \& Szeifert}]{Lemasle2014}
Lemasle, B., de~Boer, T., Hill, V., {et~al.} 2014, A\&A, 572, A88

\bibitem[{Letarte {et~al.}(2006)Letarte, Hill, Jablonka, Tolstoy,
  Fran{\c{c}}ois, \& Meylan}]{Letarte2006}
Letarte, B., Hill, V., Jablonka, P., {et~al.} 2006, A\&A, 453, 547

\bibitem[{Letarte {et~al.}(2010)Letarte, Hill, Tolstoy, Jablonka, Shetrone,
  Venn, Spite, Irwin, Battaglia, Helmi, Primas, Fran{\c{c}}ois, Kaufer,
  Szeifert, Arimoto, \& Sadakane}]{Letarte2010}
Letarte, B., Hill, V., Tolstoy, E., {et~al.} 2010, A\&A, 523, A17

\bibitem[{Luck \& Bond(1981)}]{Luck1981}
Luck, R.~E. \& Bond, H.~E. 1981, ApJ, 244, 919

\bibitem[{Martell {et~al.}(2016)Martell, Shetrone, Lucatello, Schiavon,
  Meszaros, {Allende Prieto}, {Garcia Hernandez}, Beers, \&
  Nidever}]{Martell2016}
Martell, S., Shetrone, M., Lucatello, S., {et~al.} 2016, ApJ, 825, 146

\bibitem[{Martell {et~al.}(2011)Martell, Smolinski, Beers, \&
  Grebel}]{Martell2011}
Martell, S.~L., Smolinski, J.~P., Beers, T.~C., \& Grebel, E.~K. 2011, A\&A,
  534, A136

\bibitem[{Matteucci(2001)}]{Matteucci2001}
Matteucci, F. 2001, {The chemical evolution of the Galaxy} (Dordrecht: Kluwer
  Academic Publishers)

\bibitem[{Matteucci \& Brocato(1990)}]{Matteucci1990}
Matteucci, F. \& Brocato, E. 1990, ApJ, 365, 539

\bibitem[{Matteucci \& Greggio(1986)}]{Matteucci1986}
Matteucci, F. \& Greggio, L. 1986, A\&A, 154, 279

\bibitem[{McConnachie(2012)}]{McConnachie2012}
McConnachie, A.~W. 2012, AJ, 144, 4

\bibitem[{McWilliam(1997)}]{McWilliam1997}
McWilliam, A. 1997, ARA\&A, 35, 503

\bibitem[{McWilliam(1998)}]{McWilliam1998}
McWilliam, A. 1998, AJ, 115, 1640

\bibitem[{McWilliam {et~al.}(2013)McWilliam, Wallerstein, \&
  Mottini}]{McWilliam2013}
McWilliam, A., Wallerstein, G., \& Mottini, M. 2013, ApJ, 778, 149

\bibitem[{Mucciarelli {et~al.}(2012)Mucciarelli, Bellazzini, Ibata, Merle,
  Chapman, Dalessandro, \& Sollima}]{Mucciarelli2012a}
Mucciarelli, A., Bellazzini, M., Ibata, R., {et~al.} 2012, MNRAS, 426, 2889

\bibitem[{Nissen {et~al.}(2000)Nissen, Chen, Schuster, \& Zhao}]{Nissen2000}
Nissen, P.~E., Chen, Y.~Q., Schuster, W.~J., \& Zhao, G. 2000, A\&A, 353, 722

\bibitem[{Nissen \& Schuster(1997)}]{Nissen1997}
Nissen, P.~E. \& Schuster, W.~J. 1997, A\&A, 326, 751

\bibitem[{Nissen \& Schuster(2010)}]{Nissen2010}
Nissen, P.~E. \& Schuster, W.~J. 2010, A\&A, 511, L10

\bibitem[{Pagel \& Tautvaisiene(1995)}]{Pagel1995}
Pagel, B. E.~J. \& Tautvaisiene, G. 1995, MNRAS, 276, 505

\bibitem[{Patrick {et~al.}(2015)Patrick, Evans, Davies, Kudritzki, Gazak,
  Bergemann, Plez, \& Ferguson}]{Patrick2015}
Patrick, L.~R., Evans, C.~J., Davies, B., {et~al.} 2015, ApJ, 803, 14

\bibitem[{Payne(1925)}]{Payne1925}
Payne, C.~H. 1925, PhD thesis, Harvard

\bibitem[{Peimbert \& Spinrad(1970)}]{Peimbert1970}
Peimbert, M. \& Spinrad, H. 1970, A\&A, 7, 311

\bibitem[{Pilachowski {et~al.}(1980)Pilachowski, Leep, \&
  Wallerstein}]{Pilachowski1980}
Pilachowski, C.~A., Leep, E.~M., \& Wallerstein, G. 1980, ApJ, 236, 508

\bibitem[{Pritzl {et~al.}(2005)Pritzl, Venn, \& Irwin}]{Pritzl2005}
Pritzl, B.~J., Venn, K.~A., \& Irwin, M. 2005, AJ, 130, 2140

\bibitem[{Reddy {et~al.}(2003)Reddy, Tomkin, Lambert, \& Prieto}]{Reddy2003}
Reddy, B.~E., Tomkin, J., Lambert, D.~L., \& Prieto, C.~A. 2003, MNRAS, 340,
  304

\bibitem[{Romano \& Starkenburg(2013)}]{Romano2013}
Romano, D. \& Starkenburg, E. 2013, MNRAS, 434, 471

\bibitem[{Sakari {et~al.}(2013)Sakari, Shetrone, Venn, McWilliam, \&
  Dotter}]{Sakari2013}
Sakari, C.~M., Shetrone, M., Venn, K., McWilliam, A., \& Dotter, A. 2013,
  MNRAS, 434, 358

\bibitem[{Sakari {et~al.}(2015)Sakari, Venn, Mackey, Shetrone, Dotter,
  Ferguson, \& Huxor}]{Sakari2015}
Sakari, C.~M., Venn, K.~A., Mackey, D., {et~al.} 2015, MNRAS, 448, 1314

\bibitem[{Salpeter(1955)}]{Salpeter1955}
Salpeter, E.~E. 1955, ApJ, 121, 161

\bibitem[{Sarajedini {et~al.}(2006)Sarajedini, Barker, Geisler, Harding, \&
  Schommer}]{Sarajedini2006}
Sarajedini, A., Barker, M.~K., Geisler, D., Harding, P., \& Schommer, R. 2006,
  AJ, 132, 1361

\bibitem[{Sarajedini {et~al.}(2007)Sarajedini, Bedin, Chaboyer, Dotter, Siegel,
  Anderson, Aparicio, King, Majewski, Mar{\'{i}}n-Franch, Piotto, Reid, \&
  Rosenberg}]{Sarajedini2007}
Sarajedini, A., Bedin, L.~R., Chaboyer, B., {et~al.} 2007, AJ, 133, 1658

\bibitem[{Sarajedini {et~al.}(1998)Sarajedini, Geisler, Harding, \&
  Schommer}]{Sarajedini1998}
Sarajedini, A., Geisler, D., Harding, P., \& Schommer, R. 1998, ApJ, 508, L37

\bibitem[{Sarajedini {et~al.}(2000)Sarajedini, Geisler, Schommer, \&
  Harding}]{Sarajedini2000}
Sarajedini, A., Geisler, D., Schommer, R., \& Harding, P. 2000, AJ, 120, 2437

\bibitem[{Sbordone {et~al.}(2007)Sbordone, Bonifacio, Buonanno, Marconi,
  Monaco, \& Zaggia}]{Sbordone2007}
Sbordone, L., Bonifacio, P., Buonanno, R., {et~al.} 2007, A\&A, 465, 815

\bibitem[{Schuster {et~al.}(2012)Schuster, Moreno, Nissen, \&
  Pichardo}]{Schuster2012}
Schuster, W.~J., Moreno, E., Nissen, P.~E., \& Pichardo, B. 2012, A\&A, 538,
  A21

\bibitem[{Sharina \& Davoust(2009)}]{Sharina2009}
Sharina, M. \& Davoust, E. 2009, A\&A, 497, 65

\bibitem[{Sharina {et~al.}(2010)Sharina, Chandar, Puzia, Goudfrooij, \&
  Davoust}]{Sharina2010}
Sharina, M.~E., Chandar, R., Puzia, T.~H., Goudfrooij, P., \& Davoust, E. 2010,
  MNRAS, 405, 839

\bibitem[{Shetrone {et~al.}(2003)Shetrone, Venn, Tolstoy, Primas, Hill, \&
  Kaufer}]{Shetrone2003}
Shetrone, M., Venn, K.~A., Tolstoy, E., {et~al.} 2003, AJ, 125, 684

\bibitem[{Sneden {et~al.}(1997)Sneden, Kraft, Shetrone, Smith, Langer, \&
  Prosser}]{Sneden1997}
Sneden, C., Kraft, R.~P., Shetrone, M.~D., {et~al.} 1997, AJ, 114, 1964

\bibitem[{Sneden {et~al.}(1979)Sneden, Lambert, \& Whitaker}]{Sneden1979}
Sneden, C., Lambert, D.~L., \& Whitaker, R.~W. 1979, ApJ, 234, 964

\bibitem[{Suda {et~al.}(2017)Suda, Hidaka, Aoki, Katsuta, Yamada, Fujimoto,
  Ohtani, Masuyama, Noda, \& Wada}]{Suda2017}
Suda, T., Hidaka, J., Aoki, W., {et~al.} 2017, Publications of the Astronomical
  Society of Japan, 69, 76

\bibitem[{Swan {et~al.}(2016)Swan, Cole, Tolstoy, \& Irwin}]{Swan2016}
Swan, J., Cole, A.~A., Tolstoy, E., \& Irwin, M.~J. 2016, MNRAS, 456, 4315

\bibitem[{Thomas {et~al.}(2005)Thomas, Maraston, Bender, \&
  de~Oliveira}]{Thomas2005}
Thomas, D., Maraston, C., Bender, R., \& de~Oliveira, C.~M. 2005, ApJ, 621, 673

\bibitem[{Tinsley(1979)}]{Tinsley1979}
Tinsley, B.~M. 1979, ApJ, 229, 1046

\bibitem[{Tissera {et~al.}(2012)Tissera, White, \& Scannapieco}]{Tissera2012}
Tissera, P.~B., White, S. D.~M., \& Scannapieco, C. 2012, MNRAS, 420, 255

\bibitem[{Tolstoy {et~al.}(2009)Tolstoy, Hill, \& Tosi}]{Tolstoy2009}
Tolstoy, E., Hill, V., \& Tosi, M. 2009, ARA\&A, 47, 371

\bibitem[{Trager {et~al.}(2000)Trager, Faber, Worthey, \&
  Gonz{\'{a}}lez}]{Trager2000}
Trager, S.~C., Faber, S.~M., Worthey, G., \& Gonz{\'{a}}lez, J.~J. 2000, AJ,
  120, 165

\bibitem[{Vargas {et~al.}(2014)Vargas, Geha, \& Tollerud}]{Vargas2014}
Vargas, L.~C., Geha, M.~C., \& Tollerud, E.~J. 2014, ApJ, 790, 73

\bibitem[{Vazdekis {et~al.}(2015)Vazdekis, Coelho, Cassisi, Ricciardelli,
  Falc{\'{o}}n-Barroso, S{\'{a}}nchez-Bl{\'{a}}zquez, {La Barbera}, Beasley, \&
  Pietrinferni}]{Vazdekis2015}
Vazdekis, A., Coelho, P., Cassisi, S., {et~al.} 2015, MNRAS, 449, 1177

\bibitem[{Veljanoski {et~al.}(2013)Veljanoski, Ferguson, Huxor, Mackey,
  Fishlock, Irwin, Tanvir, Chapman, Ibata, Lewis, \&
  McConnachie}]{Veljanoski2013}
Veljanoski, J., Ferguson, A. M.~N., Huxor, A.~P., {et~al.} 2013, MNRAS, 435,
  3654

\bibitem[{Veljanoski {et~al.}(2015)Veljanoski, Ferguson, Mackey, Huxor, Hurley,
  Bernard, Cote, Irwin, Martin, Burgett, Chambers, Flewelling, Kudritzki, \&
  Waters}]{Veljanoski2015}
Veljanoski, J., Ferguson, A. M.~N., Mackey, A.~D., {et~al.} 2015, MNRAS, 452,
  320

\bibitem[{Venn {et~al.}(2004)Venn, Irwin, Shetrone, Tout, Hill, \&
  Tolstoy}]{Venn2004}
Venn, K.~A., Irwin, M., Shetrone, M.~D., {et~al.} 2004, AJ, 128, 1177

\bibitem[{Venn {et~al.}(2001)Venn, Lennon, Kaufer, McCarthy, Przybilla,
  Kudritzki, Lemke, Skillman, \& Smartt}]{Venn2001}
Venn, K.~A., Lennon, D.~J., Kaufer, A., {et~al.} 2001, ApJ, 547, 765

\bibitem[{Villanova {et~al.}(2013)Villanova, Geisler, Carraro, Bidin, \&
  Munoz}]{Villanova2013}
Villanova, S., Geisler, D., Carraro, G., Bidin, C.~M., \& Munoz, C. 2013, ApJ,
  778, 186

\bibitem[{Vincenzo {et~al.}(2015)Vincenzo, Matteucci, Recchi, Calura,
  McWilliam, \& Lanfranchi}]{Vincenzo2015}
Vincenzo, F., Matteucci, F., Recchi, S., {et~al.} 2015, MNRAS, 449, 1327

\bibitem[{Vogt {et~al.}(1994)Vogt, Allen, Bigelow, Bresee, Brown, Cantrall,
  Conrad, Couture, Delaney, Epps, Hilyard, Hilyard, Horn, Jern, Kanto, Keane,
  Kibrick, Lewis, Osborne, Pardeilhan, Pfister, Ricketts, Robinson, Stover,
  Tucker, Ward, \& Wei}]{Vogt1994}
Vogt, S.~S., Allen, S.~L., Bigelow, B.~C., {et~al.} 1994, Proc. SPIE, 2198, 362

\bibitem[{Worthey {et~al.}(1992)Worthey, Faber, \& Gonzalez}]{Worthey1992}
Worthey, G., Faber, S.~M., \& Gonzalez, J.~J. 1992, ApJ, 398, 69

\end{thebibliography}

\begin{appendix}
\section{Individual abundance measurements}

\longtab[1]{
\begin{longtable}{lcc}
\caption{\label{tab:abN147HII} Individual abundance measurements for NGC~147 Hodge~II}\\
\hline\hline
Wavelength [\AA] & Value & Error \\ \hline
\endfirsthead
\caption{continued.}\\
\hline\hline
Wavelength [\AA] & Value & Error \\ \hline
\endhead
\hline
\endfoot
\mbox{[Fe/H]} \\
4209.7--4284.4 & $-1.853$ & 0.036 \\
4260.4--4336.0 & \multicolumn{2}{c}{$\ldots$} \\
4312.4--4388.9 & \multicolumn{2}{c}{$\ldots$} \\
4365.6--4443.1 & $-1.034$ & 0.041 \\
4420.1--4498.6 & $-1.663$ & 0.070 \\
4476.1--4555.6 & $-1.363$ & 0.085 \\
4533.4--4614.0 & $-1.724$ & 0.110 \\
\end{longtable}
}

\longtab[2]{
\begin{longtable}{lcc}
\caption{\label{tab:abN147HIII} Individual abundance measurements for NGC~147 Hodge~III}\\
\hline\hline
Wavelength [\AA] & Value & Error \\ \hline
\endfirsthead
\caption{continued.}\\
\hline\hline
Wavelength [\AA] & Value & Error \\ \hline
\endhead
\hline
\endfoot
\mbox{[Fe/H]} \\
4209.8--4284.4 & $-2.582$ & 0.061 \\
4260.5--4336.0 & $-2.123$ & 0.056 \\
4312.4--4388.9 & $-2.513$ & 0.106 \\
4365.7--4443.1 & $-2.502$ & 0.066 \\
4420.2--4498.7 & $-2.582$ & 0.066 \\
4476.1--4555.6 & $-2.482$ & 0.051 \\
4533.5--4614.1 & $-2.513$ & 0.085 \\
\end{longtable}
}

\longtab[3]{
\begin{longtable}{lcc}
\caption{\label{tab:abN147PA1} Individual abundance measurements for NGC~147 PA1}\\
\hline\hline
Wavelength [\AA] & Value & Error \\ \hline
\endfirsthead
\caption{continued.}\\
\hline\hline
Wavelength [\AA] & Value & Error \\ \hline
\endhead
\hline
\endfoot
\mbox{[Fe/H]} \\
4209.9--4284.6 & $-2.578$ & 0.051 \\
4260.6--4336.2 & $-1.969$ & 0.045 \\
4312.6--4389.1 & $-2.269$ & 0.071 \\
4365.8--4443.3 & $-2.218$ & 0.051 \\
4420.4--4498.9 & $-2.439$ & 0.051 \\
4476.3--4555.8 & $-2.359$ & 0.061 \\
4533.6--4614.2 & $-2.478$ & 0.096 \\
\end{longtable}
}

\longtab[4]{
\begin{longtable}{lcc}
\caption{\label{tab:abN147PA2} Individual abundance measurements for NGC~147 PA2}\\
\hline\hline
Wavelength [\AA] & Value & Error \\ \hline
\endfirsthead
\caption{continued.}\\
\hline\hline
Wavelength [\AA] & Value & Error \\ \hline
\endhead
\hline
\endfoot
\mbox{[Fe/H]} \\
4209.9--4284.6 & $-2.119$ & 0.035 \\
4260.6--4336.2 & $-1.908$ & 0.041 \\
4312.6--4389.1 & $-1.799$ & 0.046 \\
4365.8--4443.3 & $-2.170$ & 0.036 \\
4420.4--4498.9 & $-2.050$ & 0.040 \\
4476.3--4555.8 & $-1.960$ & 0.041 \\
4533.6--4614.2 & $-2.079$ & 0.051 \\
\end{longtable}
}

\longtab[5]{
\begin{longtable}{lcc}
\caption{\label{tab:abN147SD7} Individual abundance measurements for NGC~147 SD7}\\
\hline\hline
Wavelength [\AA] & Value & Error \\ \hline
\endfirsthead
\caption{continued.}\\
\hline\hline
Wavelength [\AA] & Value & Error \\ \hline
\endhead
\hline
\endfoot
\mbox{[Fe/H]} \\
4209.8--4284.4 & $-2.008$ & 0.036 \\
4260.5--4336.0 & $-2.239$ & 0.045 \\
4312.5--4388.9 & $-2.127$ & 0.060 \\
4365.7--4443.1 & $-2.048$ & 0.036 \\
4420.2--4498.7 & $-2.037$ & 0.045 \\
4476.2--4555.6 & $-2.008$ & 0.051 \\
4533.5--4614.1 & $-2.098$ & 0.051 \\
\end{longtable}
}

\longtab[6]{
\begin{longtable}{lcc}
\caption{\label{tab:abN6822SC6} Individual abundance measurements for NGC~6822 SC6}\\
\hline\hline
Wavelength [\AA] & Value & Error \\ \hline
\endfirsthead
\caption{continued.}\\
\hline\hline
Wavelength [\AA] & Value & Error \\ \hline
\endhead
\hline
\endfoot
\mbox{[Fe/H]} \\
4206.4--4280.9 & $-1.919$ & 0.035 \\
4257.0--4332.5 & $-1.839$ & 0.041 \\
4308.9--4385.4 & $-1.860$ & 0.045 \\
4362.1--4439.6 & $-1.850$ & 0.030 \\
4416.6--4495.1 & $-1.819$ & 0.041 \\
4472.5--4552.0 & $-1.749$ & 0.045 \\
4529.8--4610.3 & $-1.789$ & 0.045 \\
\end{longtable}
}

\longtab[7]{
\begin{longtable}{lcc}
\caption{\label{tab:abN6822SC7} Individual abundance measurements for NGC~6822 SC7}\\
\hline\hline
Wavelength [\AA] & Value & Error \\ \hline
\endfirsthead
\caption{continued.}\\
\hline\hline
Wavelength [\AA] & Value & Error \\ \hline
\endhead
\hline
\endfoot
\mbox{[Fe/H]} \\
4207.1--4281.6 & $-1.360$ & 0.018 \\
4257.8--4333.2 & $-1.418$ & 0.020 \\
4309.7--4386.1 & $-1.422$ & 0.025 \\
4362.9--4440.3 & $-1.330$ & 0.021 \\
4417.4--4495.8 & $-1.251$ & 0.025 \\
4473.3--4552.7 & $-1.251$ & 0.025 \\
4530.6--4611.1 & $-1.341$ & 0.030 \\
\end{longtable}
}

\longtab[8]{
\begin{longtable}{lcc}
\caption{\label{tab:abM33H38} Individual abundance measurements for M33 H38}\\
\hline\hline
Wavelength [\AA] & Value & Error \\ \hline
\endfirsthead
\caption{continued.}\\
\hline\hline
Wavelength [\AA] & Value & Error \\ \hline
\endhead
\hline
\endfoot
\mbox{[Fe/H]} \\
4216.2--4276.6 & $-1.359$ & 0.045 \\
4267.0--4328.1 & $-0.656$ & 0.040 \\
4319.0--4380.9 & \multicolumn{2}{c}{$\ldots$} \\
4372.3--4435.0 & $-0.746$ & 0.035 \\
4427.0--4490.4 & $-1.686$ & 0.057 \\
4483.0--4547.2 & \multicolumn{2}{c}{$\ldots$} \\
4599.4--4665.3 & $-1.137$ & 0.056 \\
\end{longtable}
}

\longtab[9]{
\begin{longtable}{lcc}
\caption{\label{tab:abM33M9} Individual abundance measurements for M33 M9}\\
\hline\hline
Wavelength [\AA] & Value & Error \\ \hline
\endfirsthead
\caption{continued.}\\
\hline\hline
Wavelength [\AA] & Value & Error \\ \hline
\endhead
\hline
\endfoot
\mbox{[Fe/H]} \\
4216.3--4276.7 & $-2.051$ & 0.061 \\
4267.1--4328.2 & $-1.721$ & 0.071 \\
4319.1--4381.0 & $-2.031$ & 0.090 \\
4372.4--4435.1 & $-1.630$ & 0.041 \\
4427.1--4490.5 & $-1.630$ & 0.045 \\
4483.1--4547.4 & $-1.831$ & 0.051 \\
4540.6--4605.7 & $-1.791$ & 0.060 \\
\end{longtable}
}

\longtab[10]{
\begin{longtable}{lcc}
\caption{\label{tab:abM33R12} Individual abundance measurements for M33 R12}\\
\hline\hline
Wavelength [\AA] & Value & Error \\ \hline
\endfirsthead
\caption{continued.}\\
\hline\hline
Wavelength [\AA] & Value & Error \\ \hline
\endhead
\hline
\endfoot
\mbox{[Fe/H]} \\
4215.9--4276.3 & $-1.060$ & 0.030 \\
4266.7--4327.8 & $-0.989$ & 0.036 \\
4318.7--4380.6 & $-0.840$ & 0.041 \\
4372.0--4434.6 & $-0.960$ & 0.026 \\
4426.6--4490.1 & $-0.931$ & 0.030 \\
4482.7--4546.9 & $-0.820$ & 0.030 \\
4540.1--4605.2 & $-1.040$ & 0.030 \\
\end{longtable}
}

\longtab[11]{
\begin{longtable}{lcc}
\caption{\label{tab:abM33U49} Individual abundance measurements for M33 U49}\\
\hline\hline
Wavelength [\AA] & Value & Error \\ \hline
\endfirsthead
\caption{continued.}\\
\hline\hline
Wavelength [\AA] & Value & Error \\ \hline
\endhead
\hline
\endfoot
\mbox{[Fe/H]} \\
4214.9--4275.3 & $-1.457$ & 0.051 \\
4265.7--4326.8 & $-1.908$ & 0.080 \\
4317.7--4379.6 & $-1.727$ & 0.108 \\
4371.0--4433.6 & $-1.387$ & 0.036 \\
4425.6--4489.1 & $-1.317$ & 0.056 \\
4481.6--4545.9 & $-1.422$ & 0.075 \\
4539.1--4604.1 & $-1.388$ & 0.071 \\
\end{longtable}
}

\longtab[12]{
\begin{longtable}{lcc}
\caption{\label{tab:abWLMGC} Individual abundance measurements for the WLM GC}\\
\hline\hline
Wavelength [\AA] & Value & Error \\ \hline
\endfirsthead
\caption{continued.}\\
\hline\hline
Wavelength [\AA] & Value & Error \\ \hline
\endhead
\hline
\endfoot
\mbox{[Fe/H]} \\
4200.0--4400.0 & $-1.820$ & 0.015 \\
4400.0--4600.0 & $-1.921$ & 0.015 \\
4600.0--4800.0 & $-2.001$ & 0.021 \\
4800.0--5000.0 & $-2.061$ & 0.015 \\
5000.0--5150.0 & $-1.921$ & 0.015 \\
5250.0--5400.0 & $-1.842$ & 0.026 \\
5400.0--5600.0 & $-2.061$ & 0.026 \\
\end{longtable}
}

\longtab[13]{
\begin{longtable}{lcc}
\caption{\label{tab:abF3} Individual abundance measurements for Fornax~3}\\
\hline\hline
Wavelength [\AA] & Value & Error \\ \hline
\endfirsthead
\caption{continued.}\\
\hline\hline
Wavelength [\AA] & Value & Error \\ \hline
\endhead
\hline
\endfoot
\mbox{[Fe/H]} \\
4200.0--4400.0 & $-2.391$ & 0.010 \\
4400.0--4600.0 & $-2.401$ & 0.011 \\
4600.0--4800.0 & $-2.401$ & 0.021 \\
4800.0--5000.0 & $-2.372$ & 0.011 \\
5000.0--5150.0 & $-2.272$ & 0.011 \\
5250.0--5400.0 & $-2.311$ & 0.016 \\
5400.0--5600.0 & $-2.331$ & 0.016 \\
\end{longtable}
}

\longtab[14]{
\begin{longtable}{lcc}
\caption{\label{tab:abF4} Individual abundance measurements for Fornax~4}\\
\hline\hline
Wavelength [\AA] & Value & Error \\ \hline
\endfirsthead
\caption{continued.}\\
\hline\hline
Wavelength [\AA] & Value & Error \\ \hline
\endhead
\hline
\endfoot
\mbox{[Fe/H]} \\
4200.0--4400.0 & $-1.413$ & 0.011 \\
4400.0--4600.0 & $-1.442$ & 0.005 \\
4600.0--4800.0 & $-1.382$ & 0.005 \\
4800.0--5000.0 & $-1.421$ & 0.005 \\
5000.0--5150.0 & $-1.352$ & 0.005 \\
5250.0--5400.0 & $-1.352$ & 0.011 \\
5400.0--5600.0 & $-1.352$ & 0.011 \\
\end{longtable}
}

\longtab[15]{
\begin{longtable}{lcc}
\caption{\label{tab:abF5} Individual abundance measurements for Fornax~5}\\
\hline\hline
Wavelength [\AA] & Value & Error \\ \hline
\endfirsthead
\caption{continued.}\\
\hline\hline
Wavelength [\AA] & Value & Error \\ \hline
\endhead
\hline
\endfoot
\mbox{[Fe/H]} \\
4200.0--4400.0 & $-2.071$ & 0.016 \\
4400.0--4600.0 & $-2.171$ & 0.020 \\
4600.0--4800.0 & $-2.150$ & 0.025 \\
4800.0--5000.0 & $-2.191$ & 0.021 \\
5000.0--5150.0 & $-2.171$ & 0.010 \\
5250.0--5400.0 & $-2.052$ & 0.020 \\
5400.0--5600.0 & $-2.191$ & 0.021 \\
\end{longtable}
}

\section{Abundances based on Kurucz line list}

\begin{table*}
\small
\caption{Abundance measurements.}
\label{tab:abunkur}
\centering
\begin{tabular}{l r r r r r r r r r r r}
\hline\hline
 & $\mathrm{[Fe/H]}$ & $\mathrm{[Na/Fe]}$ & $\mathrm{[Mg/Fe]}$ & $\mathrm{[Ca/Fe]}$ & $\mathrm{[Sc/Fe]}$ & 
 $\mathrm{[Ti/Fe]}$    &     $\mathrm{[Cr/Fe]}$    &     $\mathrm{[Mn/Fe]}$    &     $\mathrm{[Ni/Fe]}$ & $\mathrm{[Ba/Fe]}$ \\
 & rms$_w$ (N) & rms$_w$ (N) & rms $_w$(N) & rms$_w$ (N) & rms$_w$ (N) & rms$_w$ (N) & rms$_w$ (N) & rms$_w$ (N) & rms$_w$ (N) & rms$_w$ (N) \\
\hline
NGC 147 HII  & $-1.576$  & $+0.428$  & $+0.285$  & $+0.291$  & $+0.267$  & $+0.250$  & $-0.068$  & $-0.018$  & $+0.029$  & $-0.158$  \\
 & 0.225 (24)  & 0.086$^a$ (1)  & 0.207 (4)  & 0.195 (5)  & 0.254 (5)  & 0.266 (7)  & 0.127 (7)  & 0.298 (2)  & 0.425 (11)  & 0.391 (3) \\
NGC 147 HIII  & $-2.478$  & $+0.656$  & $+0.245$  & $+0.451$  & $+0.267$  & $+0.426$  & $-0.171$  & $+0.233$  & $-0.407$  & $-0.356$  \\
 & 0.162 (24)  & 0.090$^a$ (1)  & 0.123 (4)  & 0.055 (5)  & 0.375 (5)  & 0.125 (7)  & 0.288 (8)  & 0.278 (2)  & 0.636 (6)  & 0.121 (4) \\
NGC 147 PA-1  & $-2.379$  & $+0.577$  & $+0.346$  & $+0.264$  & $+0.062$  & $+0.380$  & $-0.010$  & $+0.067$  & $+0.070$  & $+0.114$  \\
 & 0.127 (23)  & 0.150$^a$ (1)  & 0.127 (5)  & 0.217 (8)  & 0.362 (5)  & 0.168 (9)  & 0.331 (10)  & 0.080 (1)  & 0.106 (4)  & 0.187 (4) \\
\hline
\end{tabular}
\tablefoot{
For each cluster, the first line gives the average abundance ratio and the second line gives the weighted rms and number of individual measurements.
(a): For $\mathrm{[Na/Fe]}$ we list the errors on the individual measurements of the doublet at 5683/5688 \AA .
}
\end{table*}

\end{appendix}

\end{document}